\documentclass[prl,preprint,floatfix,superscriptaddress,showpacs,nofootinbib]{revtex4-1}
\usepackage[colorlinks,linkcolor=red,citecolor=blue,urlcolor=blue]{hyperref}
\usepackage{dcolumn}
\usepackage{textcomp}
\usepackage[pdftex]{graphicx}
\usepackage{latexsym,amsmath,amssymb}
\usepackage[english]{babel}
\usepackage[pdftex]{color}
\usepackage{microtype}
\usepackage{comment}
\usepackage[normalem]{ulem}
\usepackage[final]{pdfpages}

\begin{document}
\title{Solving the Dynamic Correlation Problem of the Susceptible-Infected-Susceptible Model on Networks}
\author{Chao-Ran Cai}
\author{Zhi-Xi Wu}\email[Corresponding author: ]{wuzhx@lzu.edu.cn}
\affiliation{Institute of Computational Physics and Complex Systems, Lanzhou University, Lanzhou, Gansu 730000, China}
\author{Michael Z.~Q. Chen}
\affiliation{Department of Mechanical Engineering, The University of Hong Kong, Pokfulam Road, Hong Kong}
\author{Petter Holme}
\affiliation{Department of Energy Science, Sungkyunkwan University, Suwon 440-746, Korea}
\author{Jian-Yue Guan}
\affiliation{Institute of Computational Physics and Complex Systems, Lanzhou University, Lanzhou, Gansu 730000, China}

\begin{abstract}
The susceptible-infected-susceptible (SIS) model is a canonical model for emerging disease outbreaks. Such outbreaks are naturally modeled as taking place on networks. A theoretical challenge in network epidemiology is the dynamic correlations coming from that if one node is infected, then its neighbors are likely to be infected. By combining two theoretical approaches--the heterogeneous mean-field theory and the effective degree method--we are able to include these correlations in an analytical solution of the SIS model. We derive accurate expressions for the average prevalence (fraction of infected) and epidemic threshold. We also discuss how to generalize the approach to a larger class of stochastic population models.
\end{abstract}
\maketitle

The susceptible-infected-susceptible (SIS) model is a fundamental model of outbreaks of diseases  (like influenza, chlamydia, gonhorrea, etc.)\ that does not give immunity upon recovery. Diseases spread over networks of people and the structure of these networks affect spreading~\cite{keeling_rev}; thus, it makes sense to put the SIS model on networks. The SIS model divides the population into two classes--susceptible (S) and infected (I). Links between S and I transmit the disease (i.e.\ make the susceptible infected) with a rate $r$. Infected individuals become susceptible again with a rate $g$. One can reduce these two parameters to one--$\lambda=r/g$--the rate of new infections per SI link per recovery, or the \textit{effective infection rate}. In the thermodynamic limit $N\rightarrow\infty$, there can be a threshold, or phase-transition phenomenon when tuning $\lambda$~\cite{Marrobook}. For $\lambda$ less than a critical $\lambda_c$ the disease dies out spontaneously. If $\lambda>\lambda_c$ the disease will live forever--i.e.\ it has reached an \textit{endemic} state where the \textit{prevalence} $\rho$ (fraction of infected nodes) is nonzero. Also, for finite networks, there is effectively an endemic state as the expected extinction time, even for small networks, grows extremely fast beyond $\lambda_c$~\cite{holme_sis}. The two main directions in the literature are to study extinction times in finite, homogeneous networks~\cite{nasell} or how the threshold depends on the network structure~\cite{hmf_1,hmf_3}. This work belongs to the latter class.

For the SIS model in a well-mixed population, the threshold happens at $\lambda_c=1$~\cite{hethcoterev}. For the SIS model in homogeneous networks, such as the Erd\H{o}s-R\'{e}nyi random networks and random regular graphs, a simple approximative (mean-field) analysis gives the threshold $\lambda_c=1/\langle k\rangle$~\cite{hmf_1}. For the SIS model in scale-free networks~\cite{rmp2002}--where the degree distribution (the probability that a random node is connected to $k$ other nodes) follows $P(k)\sim k^{-\gamma}$--heterogeneous mean-field theory predicts that the epidemic threshold of the SIS model is equal to that of the susceptible-infected-recovered (SIR) model $\lambda_c^{\mathrm{HMF,SIS}}=\lambda_c^{\mathrm{HMF,SIR}}=\langle k\rangle/\langle k^2\rangle$~\cite{hmf_1,hmf_2,hmf_3,th}. In other words, the threshold seems to be zero for $\gamma\leq3$, and finite for $\gamma>3$~\cite{ss_2}. The heterogeneous mean-field theory is a degree-based mean-field theory, which sorts the nodes into different classes in terms of the magnitude of their degrees, but all the other aspects are considered totally random (for instance, who connects whom). In other words, the heterogeneous mean-field theory neglects correlation, both from the disease dynamics (a node is more likely infected if its neighbors are infected) and the network structure. Effectively, the heterogeneous mean-field theory applies to a situation where the network is constantly rewired (or \textit{annealed}) at a time scale faster than the disease dynamics. (See the Supplemental Material~\cite{sm} for more details on the SIS and SIR models on annealed networks)

Moving beyond the heterogeneous mean-field assumption that the network is rapidly changing, we have to deal with dynamic correlations. In the heterogeneous mean-field, the probability to be infected is $\rho$ for two nodes of the same degree, but because of dynamic correlations a node is more likely to be infected if many of its neighbors are infected. This fact--the probability of having an already infected neighbor is larger than in the mean-field approximation--will reduce the effective infection rate, and thus the speed and extent of the disease propagation. One possible remedy is to consider an individual-based mean-field approximation by taking into account the full network structure correlation (still ignoring the dynamic correlations). This is also called the quenched mean-field theory~\cite{qmf_1, qmf_2}. The quenched mean-field theory gives the epidemic threshold  $\lambda_c^{\mathrm{QMF,SIS}}=1/\Lambda_1$~\cite{qmf_1}, where $\Lambda_1$ is the largest eigenvalue of the adjacency matrix of the underlying network. According to Ref.~\cite{Chung27052003}, for scale-free networks,
\begin{equation}
1/\Lambda_1\sim\left\{
    \begin{array}{cl}
    1/\sqrt{k_{\mathrm{max}}}, & \gamma>5/2,\\
    \langle k\rangle/\langle k^2\rangle, & 2<\gamma<5/2,\\
    \end{array}
    \right.
\end{equation}
where $k_{\mathrm{max}}$ is the maximum degree in the network. Both the heterogeneous mean-field and quenched mean-field methods imply that the epidemic threshold in scale-free networks vanishes in the thermodynamic limit, but remains finite for networks of finite sizes. Some numerical studies suggest that the quenched mean-field is qualitatively correct in scale-free networks~\cite{Claudio2010prl,PhysRevLett.111.068701}, but others point at deviations from the threshold value~\cite{th}.

There are studies that do take dynamic correlations into account. This, so called, effective degree approach~\cite{ed_1, ed_2} is a higher order degree-based mean-field theory, explicitly considering the dynamic correlation between directly connected neighbors. It does indeed provides a much more accurate prediction of the epidemic threshold for the SIR model in uncorrelated networks, $\lambda_c^{\mathrm{ED,SIR}}=\frac{\langle k\rangle}{\langle k^2\rangle-2\langle k\rangle}$~\cite{ed_1}. This works by mapping the SIR process to a bond-percolation problem~\cite{ss_2,Grassberger}, but has not yet been applied to the SIS model. Bogu\~{n}\'{a} \emph{et al.}~\cite{PhysRevLett.111.068701} take dynamic correlation between distant neighbors into account within the framework of quenched mean-field theory. They replaced the original SIS dynamics by a modified process valid over coarse-grained times and argued that dynamic correlation changes the dynamics near the threshold.
Based on a spectral approach, which takes into account the other eigenvectors of the adjacency matrix than $\Lambda_1$, Goltsev \emph{et al.}~\cite{PhysRevLett.109.128702} showed that in scale-free networks with $\gamma>5/2$, the principal eigenvector is localized when the effective infection rate is slightly above $\lambda_c^{\mathrm{QMF, SIS}}$. Since in quenched mean field theory the density of infected vertices is proportional to the principal eigenvector, their work predicts a transition to a localized phase, where the activity is concentrated to the hubs and their immediate neighbors. Ferreira \emph{et al.}~\cite{Ferreira2016pre} studied the relationship between the hub lifespan and the hub infection time to discern the nature of the threshold in general epidemic models on scale-free networks. Reference~\cite{ruhi} presents yet further refinements of the quenched mean-field theory. Finally, in this literature review, we also want to mention many other approaches to understand SIS or SIR processes on networks, such as: the fluctuation theory~\cite{PhysRevE.87.062812}, the pair-approximation method~\cite{pa_1, pa_2}, probability generating function techniques~\cite{sn_1, sn_2}, $R_0$-based modification~\cite{sn_3}, percolation theory~\cite{sn_1, PhysRevLett.104.258701}, and branching process~\cite{NC}, etc. However, all these methods depend on large sets of coupled ordinary differential equations, and are unable to provide explicit analytical solutions for the prevalence, in particular, for large-scale networks~\cite{ss_2}.

Our current understanding of the SIS dynamics on heterogeneous networks is far from complete. In this Letter, we combine the idea of the heterogeneous mean-field theory with the effective degree approach to study the effect of dynamic correlation present in static networks on the SIS epidemic dynamics. The underlying static networks can have arbitrary degree distributions. However, we focus on networks without degree-degree correlations or other (e.g.\ mesoscopic) structures. Our method gives a closed-form analytical solution for the epidemic prevalence $\rho$, from which we immediately can obtain the epidemic threshold by solving the equation $\rho=0$. We further show that our method matches numerical simulations better than the above-mentioned theory.

To begin, we define $p_k$ and $q_k$, respectively, as the probabilities of reaching an arbitrary infected individual by following a randomly chosen edge from susceptible and infected individuals of degree $k$. In the framework of heterogeneous mean-field theory where the underlying network is treated as annealed, the value of $q_k$ is always equal to that of $p_k$~\cite{PhysRevLett.111.068701}. For the earlier mentioned reasons, the  dynamic correlations mean that
\begin{equation}\label{eq.1}
q_k>p_k .
\end{equation}
In the spirit of the heterogeneous mean-field~\cite{hmf_1}, we can write down the master equation for those infected individuals of degree class $k$,
\begin{equation}\label{eq:mstr}
\frac{dI_k(t)}{dt}=-I_k(t)g+rkS_k(t)p_k(t),
\end{equation}
where $S_k(t)$ and $I_k(t)$ are the number of susceptible and infected individuals with degree $k$ at time $t$, respectively. The first term represents the spontaneous recovery and the
second one the newly emerged infection in class $k$ due to the interaction with other classes. In the steady state, we have
\begin{equation}
\label{eq.2}
p_k=\frac{g}{kr}\frac{I_k}{S_k}.
\end{equation}
In annealed networks, one assumes $p_k=q_k=\frac{1}{\langle k\rangle N}\sum_kkI_k$, where $N$ is the total population size, to continue the theoretical derivation. As mentioned above, this hypothesis is not applicable for the SIS model in static networks. In what follows, we circumvent this problem following the effective degree approach.
This means that we divide the network into classes representing both the state of an individual and its neighbors~\cite{Gleeson_1,Gleeson_2}. Let $S_{k,j}(t)$ $[I_{k,j}(t)]$ be the density of $k$ degree nodes that are susceptible (infected) at time $t$, connected to $j$ infected neighbors. Then $S_{k,j}(t)$ and $I_{k,j}(t)$ can be written as
\begin{subequations}\label{eq:master}
\begin{eqnarray}
S_{k,j}(t)&=&S_k(t)\left[1-p_k(t)\right]^{k-j}
\left[p_k(t)\right]^j{k\choose j}, \\
I_{k,j}(t)&=&I_k(t)\left[1-q_k(t)\right]^{k-j}
\left[q_k(t)\right]^j{k\choose j},
\end{eqnarray}
\end{subequations}
where $S_k(t)=\sum_jS_{k,j}(t)$ and $I_k(t)=\sum_jI_{k,j}(t)$. Summing over all possible events, we obtain the total recovery rate $a(t)=\sum_k\sum_jI_{k,j}(t)g$ and the population-level transmission rate $b(t)=\sum_k\sum_jS_{k,j}(t)jr$. When the system is in its steady state, the total recovery rate must be equal to the total transmission rate, satisfying the \textit{detailed balance} conditions
$a=b$ and $\langle\Delta a\rangle=\langle\Delta b\rangle$,
where $\Delta a$ and $\Delta b$ are the changing rate of $a$ and $b$ in the time interval $dt$ (where only one event occurs). All possible values of these two quantities in a static network are summarized in Table~\ref{ta.1}. With these preliminary results, we have
\begin{equation}
\label{eq.3}
\sum_k\sum_jS_{k,j}j^2r=\sum_k\sum_jI_{k,j}jg.
\end{equation}
Note that the formula of Eq.~(\ref{eq.3}) is exact for the steady state in any static networks. Nevertheless, we want to point out that they are not applicable to irreversible processes such as the SIR model.

\begin{table}
\caption{All possible situations in a time interval  $dt$ of the SIS process in a static network.\label{ta.1}}
\begin{ruledtabular}
\begin{tabular}{cccc}
{Situation}&{Probability}&{$\Delta a$}&{$\Delta b$}\\
\colrule
$S_{k,j}\rightarrow I_{k,j}$ & $\frac{S_{k,j}(t)jr}{a+b}$ & $+g$ & $-jr+r(k-j)$\\
$I_{k,j}\rightarrow S_{k,j}$ & $\frac{I_{k,j}(t)g}{a+b}$ & $-g$ & $+jr-r(k-j)$\\
\end{tabular}
\end{ruledtabular}
\end{table}

Using the condition that the total number of infectious neighbors of all susceptible individuals equals the total number of susceptible neighbors of all infectious individuals~\cite{keeling}, we get the relationship $\sum_k\sum_jI_{k,j}jg=g\sum_kI_k[k-g/r]$. Combining this with Eqs.~(\ref{eq.2}) and (\ref{eq.3}), gives
\begin{equation}
\label{eq.4}
g\sum_kI_k\left[(k-1)p_k+1-k+\frac{g}{r}\right]=0.
\end{equation}

\begin{figure}
\includegraphics[width=0.49\linewidth]{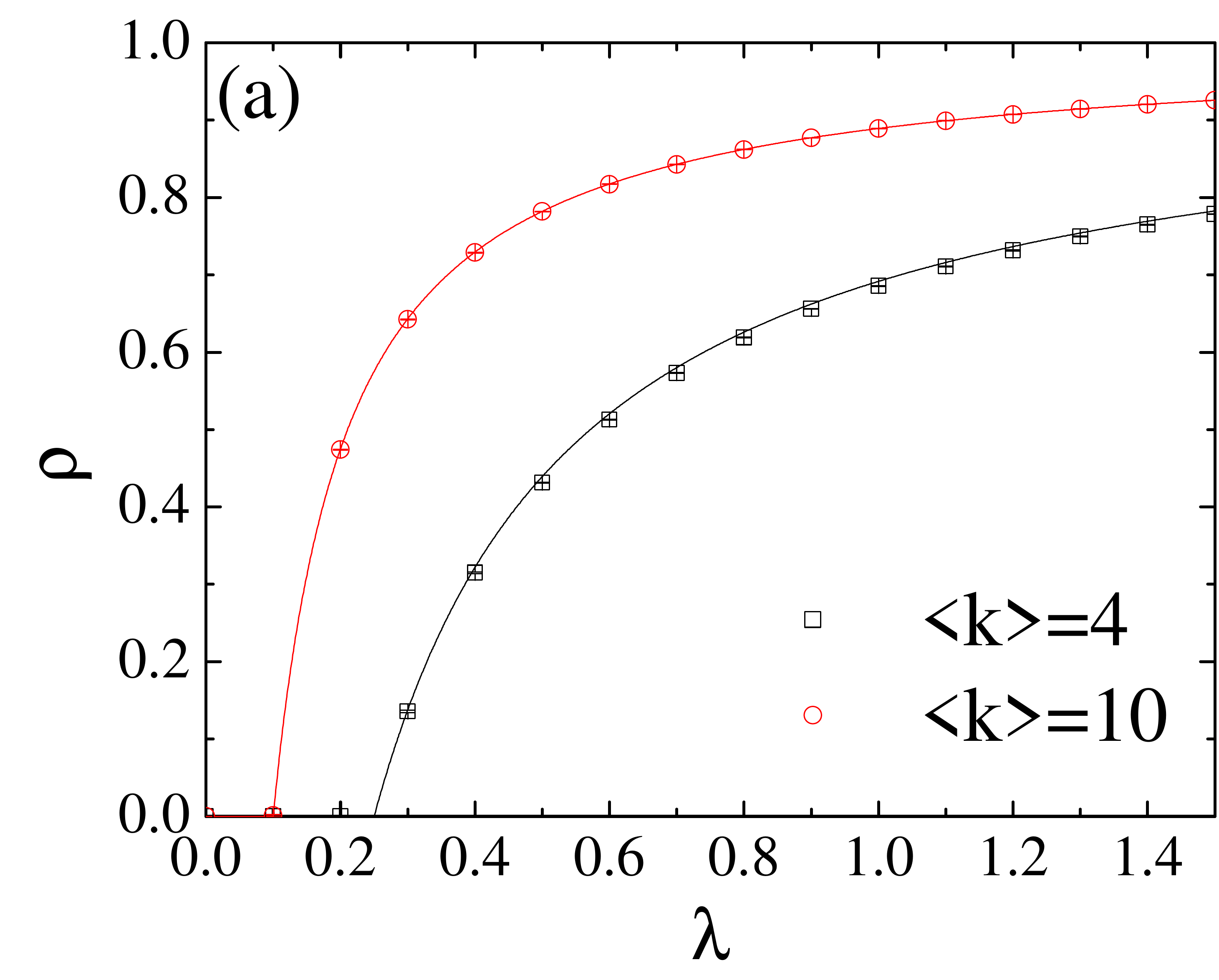}
\includegraphics[width=0.49\linewidth]{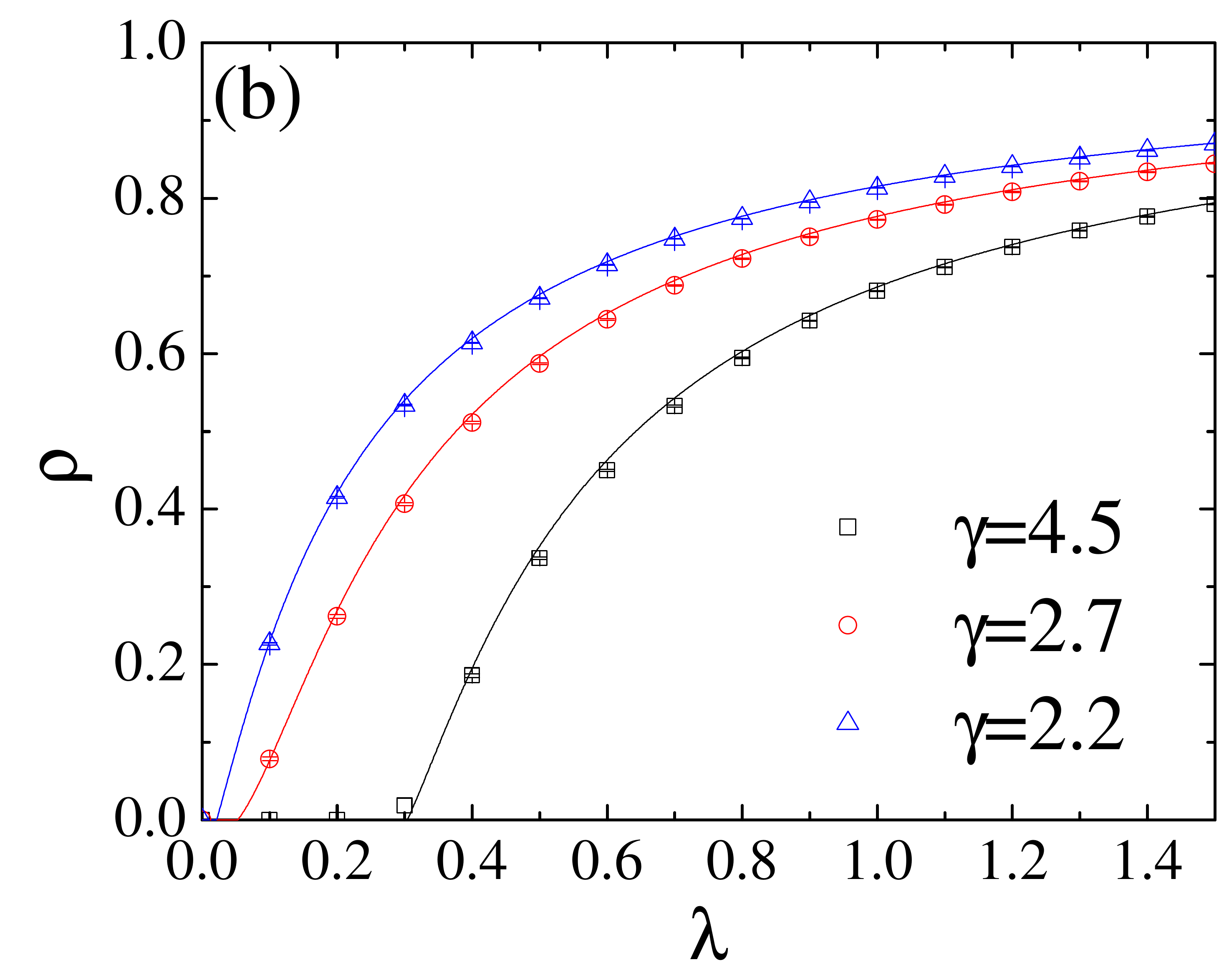}\\
\caption{The epidemic prevalence $\rho$ is plotted as a function of the effective infection rate $\lambda$ in static networks. (a) Erd\H{o}s-R\'{e}nyi random networks with average degree $4$ and $10$, (b) scale-free network with minimum degree $3$ and $\gamma=4.5$, $2.7$, and $2.2$. Lines are theoretical estimations of Eqs.~(\ref{eq.6}) and (\ref{eq.8}), while the points are simulation results. We use networks with $N=10^5$ nodes and $100$ randomly chosen seeds for the infection. Each data point is an average over at least $100$ independent epidemic outbreaks, performed on at least $10$ different network realizations.}\label{fig.1}
\end{figure}

Notice that only infected individuals can determine the birth and death of susceptible individuals, but not vice versa. Any new infection or recovery event will mainly change the difference among various $q_k$ (since the probability of finding connected infected pair with certain degrees will increase or decrease definitely due to the event). But it does not provide any valuable hints about how various $p_k$ will differ from each other (albeit their values would change owing to the newly emerged or disappeared infected individual) for \textit{degree-uncorrelated} random networks. In light of the maximum entropy principle~\cite{Jaynes1957}, we assume all $p_k$ change with the same magnitude such that they always satisfy the following approximative relations
\begin{equation}
\label{eq.5}
p_1\simeq p_2\simeq \cdots \simeq p_{k_{\mathrm{max}}}=p.
\end{equation}
That is to say, our method is mainly concerned about the dynamic correlations present in connected infected pairs, but neglects higher-order dynamic correlations in susceptible-susceptible or susceptible-infected pairs. We have verified that Eqs.~(\ref{eq.1}) and (\ref{eq.5}) capture the most important parts of the dynamic correlations, neglecting only minor ones~\cite{sm}. Substituting Eq.~(\ref{eq.5}) into Eqs.~(\ref{eq.2}) and (\ref{eq.4}), we obtain the iterative equation
\begin{equation}
\label{eq.6}
I_k=\frac{\lambda kp}{1+\lambda kp}N_k,
\end{equation}
where $p$ itself is a function of $I_k$ as
\begin{equation}
\label{eq.7}
p=1-\frac{I}{\lambda \sum_kI_k(k-1)}.
\end{equation}
Combining with Eqs.~(\ref{eq.6}) and (\ref{eq.7}), we can find the solution of $p$, which is now a function of $\lambda$, satisfying the self-consistency equation
\begin{equation}
\label{eq.8}
(1-p)\lambda\sum_k(k-1)\frac{\lambda kp}{1+\lambda kp}NP(k)-\sum_k\frac{\lambda kp}{1+\lambda kp}NP(k)=0.
\end{equation}

Now we are able to calculate the epidemic prevalence $\rho$ in the steady state as follows: (i) Calculate $p$ from Eq.~(\ref{eq.8}); (ii) Substitute the value of $p$ into Eq.~(\ref{eq.6}) to solve $I_k$; (iii) Obtain the epidemic prevalence $\rho=\frac{1}{N}\sum_kI_k$. The results are summarized in Fig.~\ref{fig.1}, from which we can see that the estimations obtained by our approach match those from stochastic simulations quite well~\cite{sm}.

A nonzero stationary epidemic prevalence is obtained when the $p$ has a nontrivial solution in the interval $0<p\leq1$. We denote the left-hand side of Eq.~(\ref{eq.8}) by $f(p)$. It is easy to see that $p=0$ is a trivial solution of Eq.~(\ref{eq.8}). Furthermore, note that $f(p)$ is always negative for $p=1$. Hence, the condition that $p$ has a meaningful solution in the interval $(0,1]$ reads as 
\begin{equation}
\frac{d}{dp}f(p)\bigg|_{p=0}\geq0.
\end{equation}
The value of $\lambda$ satisfying the equality of the above inequality determines the epidemic threshold $\lambda_c$, whose value is given, for uncorrelated random networks, by
\begin{equation}
\label{eq.9}
\lambda_c=\frac{\langle k\rangle}{\langle k^2\rangle-\langle k\rangle}.
\end{equation}

\begin{figure}
\includegraphics[width=0.49\linewidth]{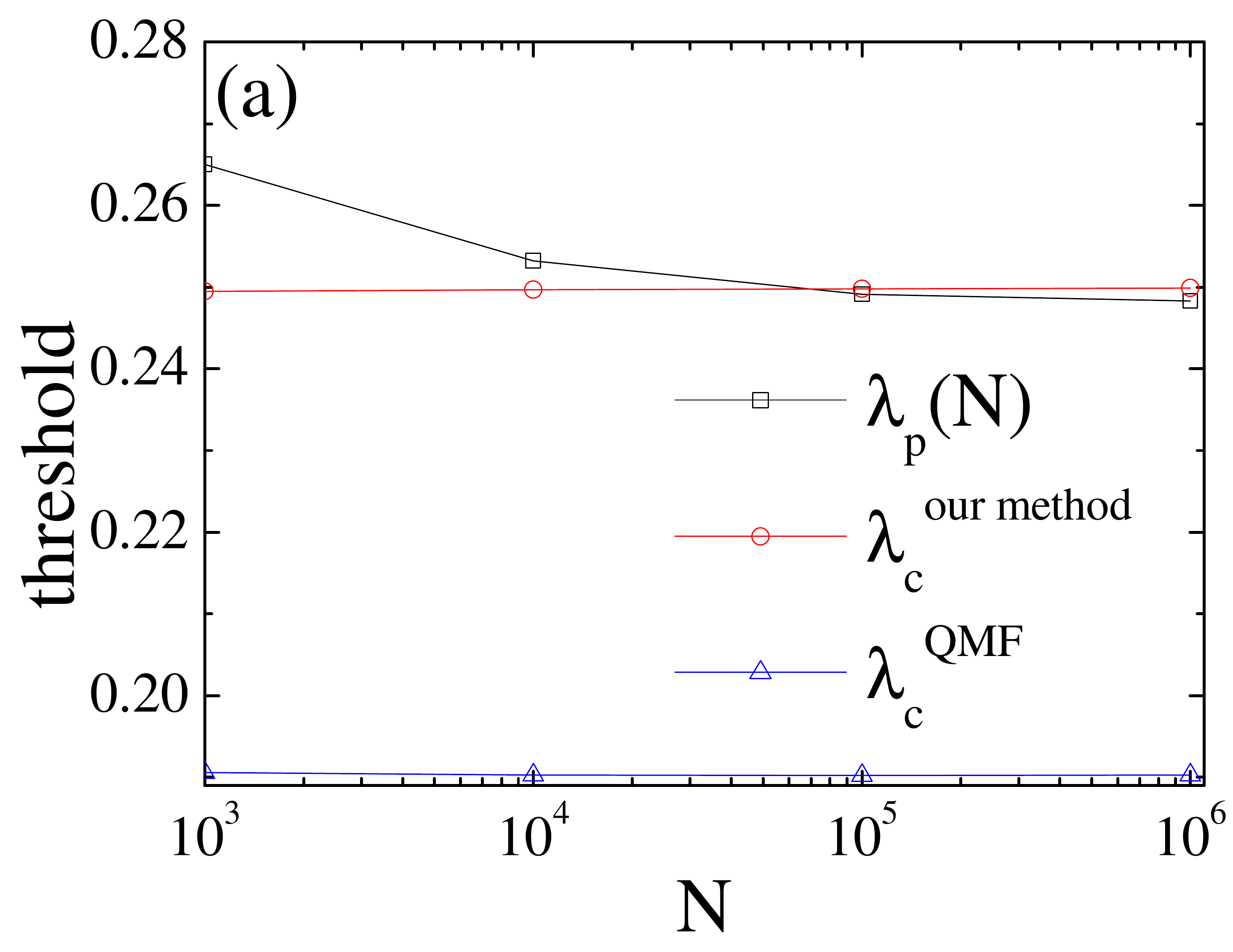}
\includegraphics[width=0.49\linewidth]{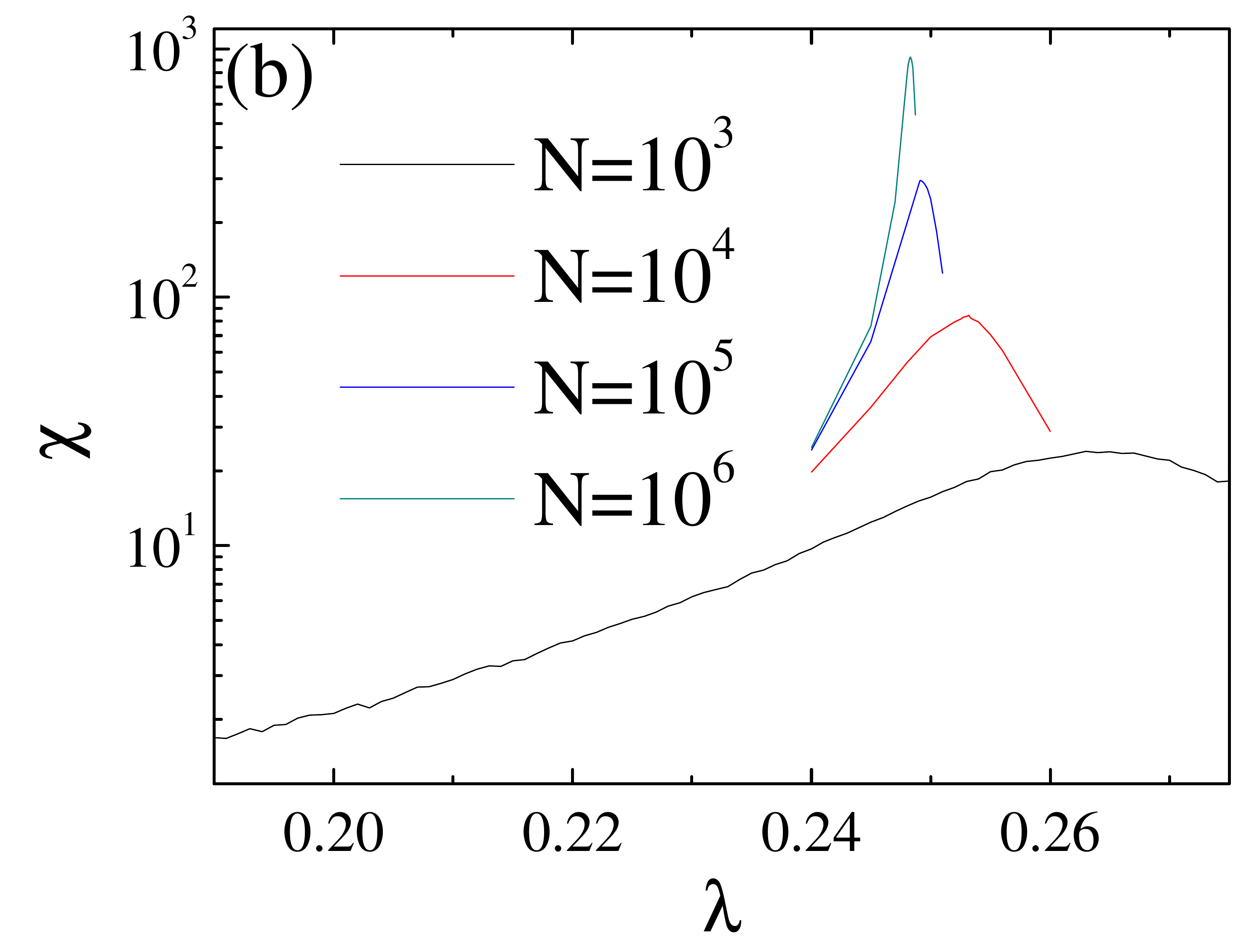}
\includegraphics[width=0.49\linewidth]{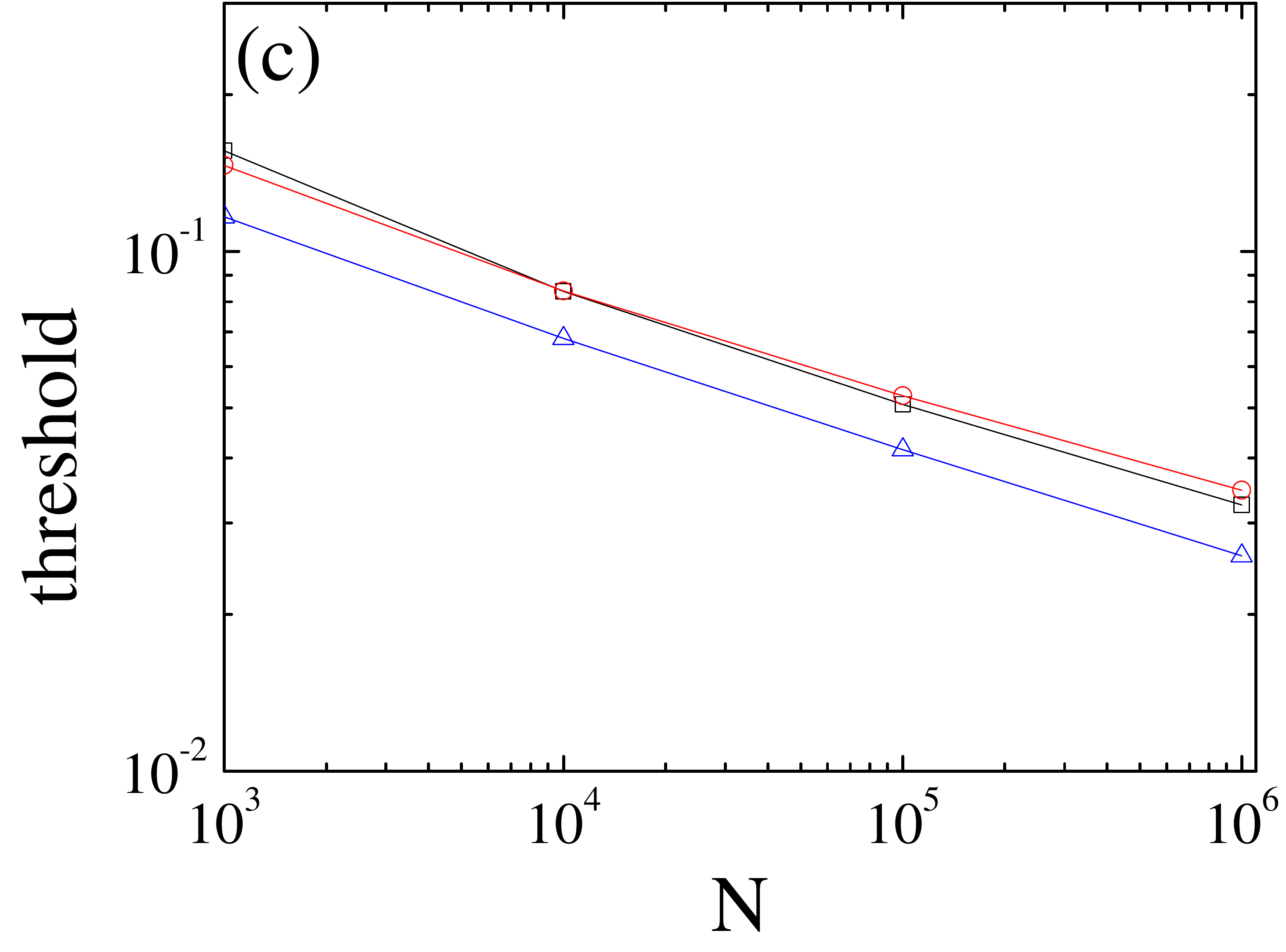}
\includegraphics[width=0.49\linewidth]{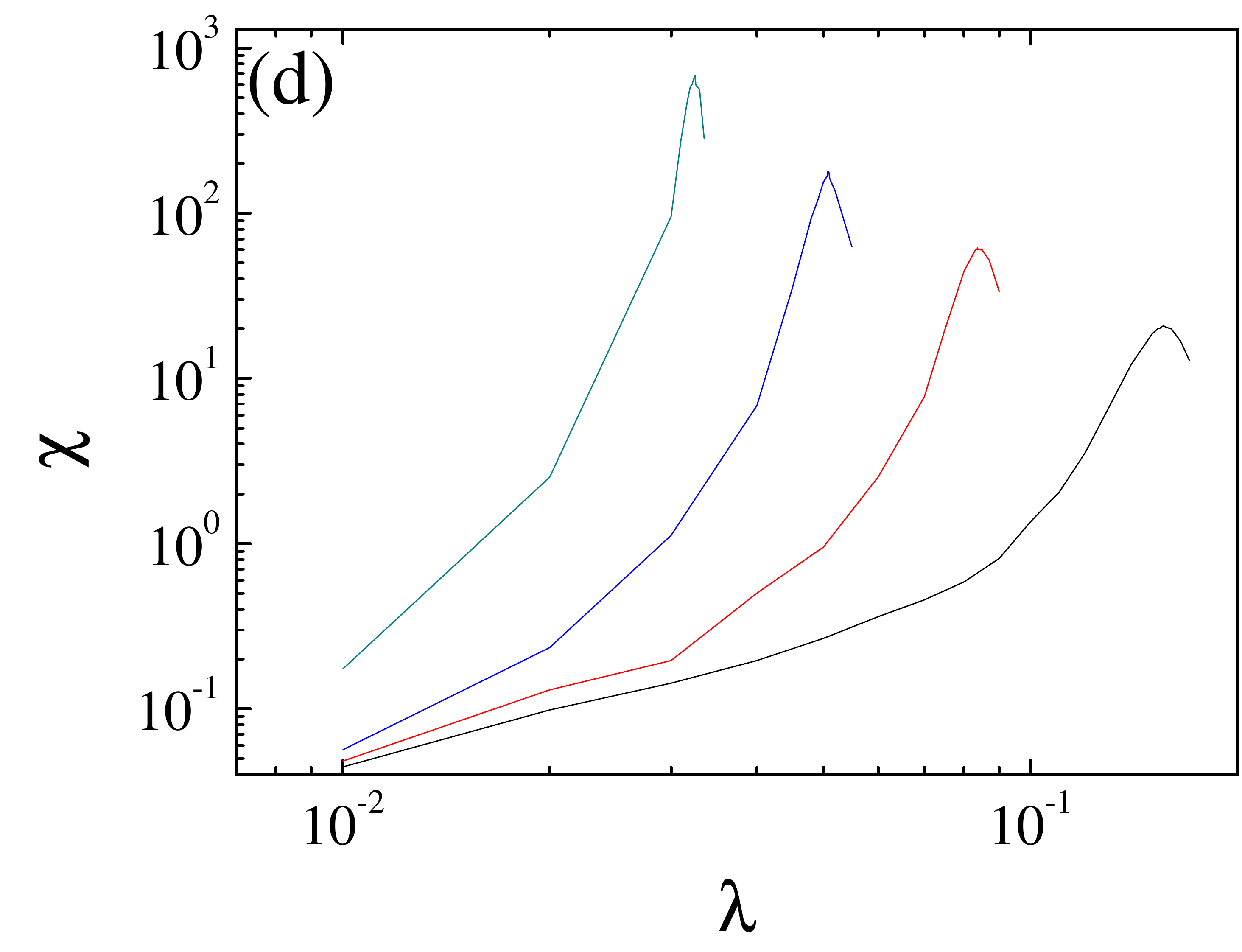}\\
\caption{(a) and (c) Epidemic threshold $\lambda_c$, obtained by  our method, the quenched mean-field method, and the quasistationary state method~\cite{th}, versus network size $N$. (b) and (d) Susceptibility $\chi$ as a function of the effective infection rate $\lambda$ for varying network size $N$, where the peak values determine the $\lambda_p(N)$ in (a) and (c). Panels (a) and (b) are for Erd\H{o}s-R\'{e}nyi random networks with average degree $\langle k\rangle=4$; (c) and (d) are for scale-free networks with minimum degree $3$ and $\gamma=2.7$. The results are obtained for the epidemic dynamics over at least ten different network realizations with initial infected seed number $I_0=1$. }\label{fig.2}
\end{figure}

The epidemic threshold of the SIS model obtained by our method is larger than the one predicted by the heterogeneous mean-field theory~\cite{hmf_1, hmf_2, hmf_3, th}, and is smaller than the threshold  of the SIR model by the effective degree approach~\cite{ed_1, ss_2}. Equation~(\ref{eq.9}) implies that the epidemic threshold of the SIS epidemic process is zero  for scale-free networks with $\gamma<3$, but finite if $\gamma>3$. This is consistent with the conclusions of previous work using other methods~\cite{PhysRevLett.109.128702,PhysRevE.87.062812}. In Fig.~\ref{fig.2}, we plot the epidemic thresholds against network size and show that, in Erd\H{o}s-R\'{e}nyi and scale-free random networks, the accuracy of the epidemic threshold of our method is better than that of the quenched mean-field theory, and matches  the results from the quasistationary numerical simulation well~\cite{th}.

\begin{figure}
\includegraphics[width=0.49\linewidth]{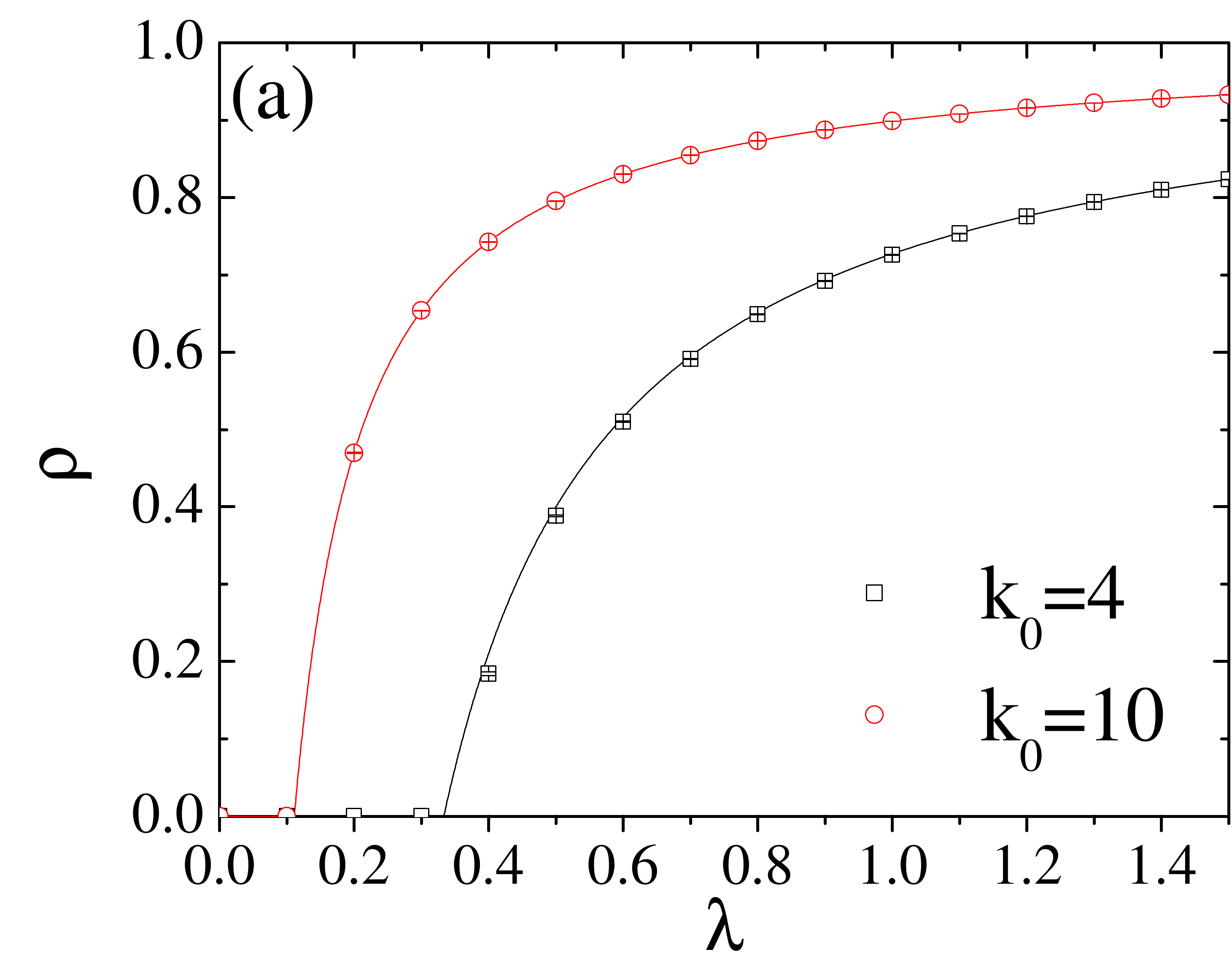}
\includegraphics[width=0.49\linewidth]{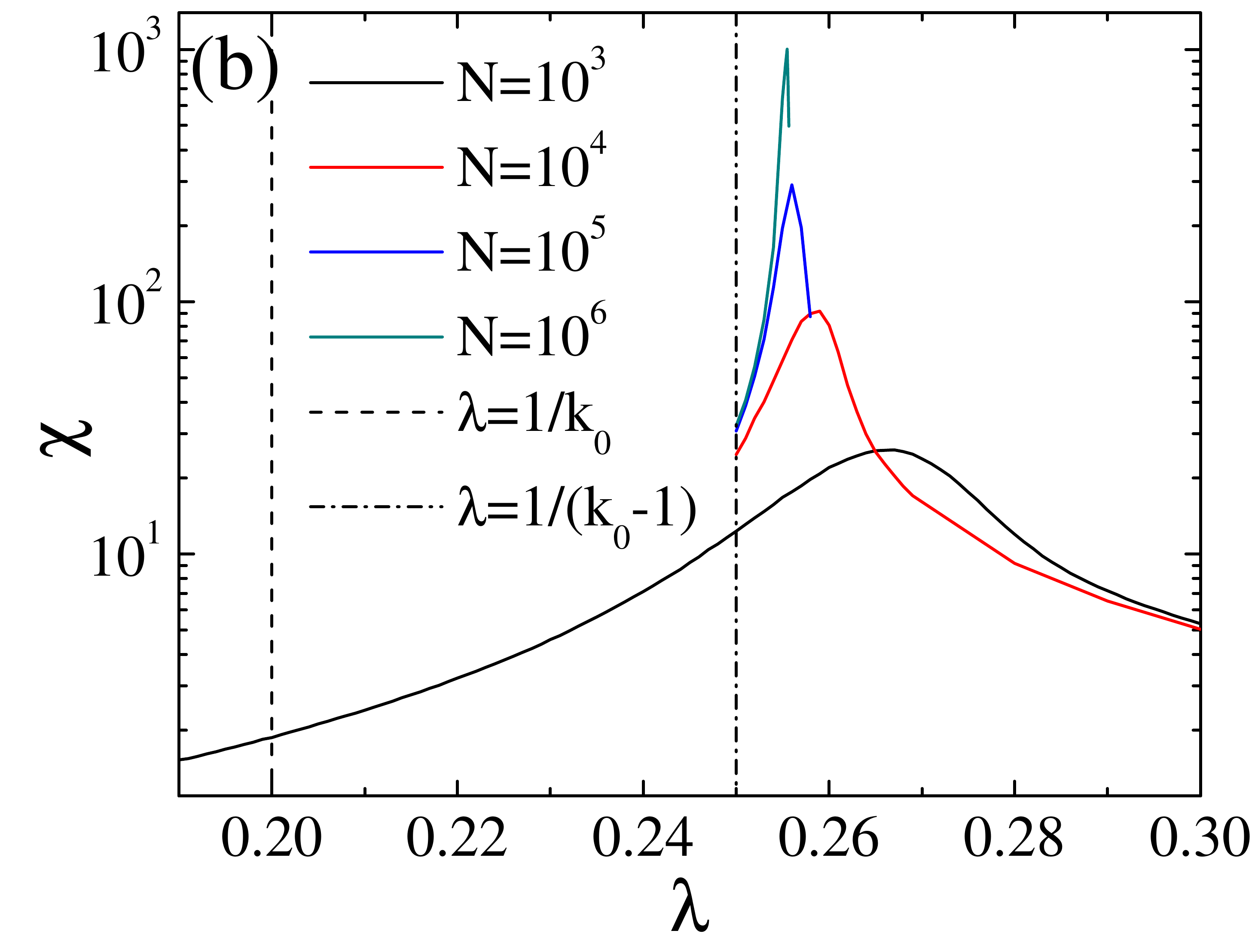}\\
\caption{(a) The epidemic prevalence $\rho$ is plotted as a function of the effective infection rate $\lambda$ in static random regular graphs with degree $k_0=4$ and $10$. Lines are the theoretical predictions of Eq.~(\ref{eq.10}), while the scatters are Monte Carlo simulation results. Other parameters are the same as those in Fig.~\ref{fig.1}. (b) Susceptibility $\chi$ as a function of the effective infection rate $\lambda$ for random regular graphs with different network sizes, where $k_0$ is fixed to 5.}\label{fig.4}
\end{figure}

To provide further evidence on the efficiency of our proposed approach, we consider the SIS model in a random regular graph whose degree distribution is a Kronecker's delta function $P(k)=\delta_{kk_0}$; i.e., each node in a random regular graph has a degree of $k_0$ and all the other aspects are totally random. From Eqs.~(\ref{eq.2}) and (\ref{eq.4}), we can acquire the epidemic prevalence
\begin{equation}
\label{eq.10}
\rho=1-\frac{1}{1+k_0(\lambda-\frac{1}{k_0-1})},
\end{equation}
where $\lambda=r/g$ is again the effective infection rate. From Fig.~\ref{fig.4}(a), we can see that the results from the analytical solution of Eq.~(\ref{eq.10}) are in excellent agreement with those obtained from stochastic simulations in a static random regular graph.

For the epidemic threshold of the SIS model in static random regular graphs, the prediction of the quenched mean-field theory can easily be simplified to $\lambda_c^\mathrm{QMF}=1/k_0$ by applying the Perron-Frobenius theorem~\cite{th}. By means of the pair-approximation method, a more accurate estimation of epidemic threshold, $\lambda_c^{\text{pair}}=1/(k_0-1)$, is reported~\cite{pa_1}. Very recently, by combining the branching process with the probability generating function, Leventhal \emph{et al.} obtained the same threshold $1/(k_0-1)$~\cite{NC}. For our case, we just need to set $\rho=0$ in Eq.~(\ref{eq.10}), and then the epidemic threshold can be obtained directly as
\begin{equation}
\label{eq.11}
\lambda_c=\frac{1}{k_0-1},
\end{equation}
which is consistent with the findings in Refs.~\cite{pa_1,NC}. In Fig.~\ref{fig.4}(b), we plot the susceptibility against the effective infection rate $\lambda$ in static random regular graph with degree $k_0=5$, and show that the susceptibility peak is closer to the theoretical prediction of Eq.~(\ref{eq.11}) than to the quenched mean-field result~\cite{th}, once again validating our method.

Thus, the combination of the heterogeneous mean-field theory and the effective degree method enables us to, on one hand, include dynamic correlation and network-structural correlation (to a necessary extent) to obtain more accurate predictions (than in previous works) of both the epidemic prevalence and threshold, and on the other hand obtain explicit expressions for these two quantities.

As a final analysis, we extended our method to the case of contact process~\cite{PhysRevLett.96.038701}--where infected individuals meet random neighbors for possible contagion events--on random regular graph networks with degree $k_0$. According to the heterogeneous mean-field theory, we have $p=\frac{g}{r}\frac{I_{k_0}}{S_{k_0}}$. The total recovery and transmission rates of the whole population are $a=\sum_jI_{k_0,j}g$ and $b=\sum_jS_{k_0,j}jr\frac{1}{k_0}$, respectively. Equation~(\ref{eq.3}) will now be replaced by $\frac{r}{k_0}\sum_jS_{k_0,j}j^2=g\sum_jI_{k_0,j}j$. Then, we can straightforwardly derive the  prevalence as
\begin{equation}
\label{eq.12}
\rho=1-\frac{1}{1+\lambda-\frac{k_0}{k_0-1}}.
\end{equation}
Accordingly, the epidemic threshold of the contact process on random regular graphs is given by
\begin{equation}
\label{eq.13}
\lambda_c=\frac{k_0}{k_0-1}.
\end{equation}
This epidemic threshold is also consistent with previously reported results~\cite{PhysRevE.84.066102,cp2010prl}; see more details in~\cite{sm}.

In summary, we have studied the impact of dynamic correlations, naturally arising in spreading processes on static networks, on the SIS epidemics. In particular, we take into account the dynamic correlation from infected pairs, but ignore those from other node pairs and higher-order network structure, to derive the master equations governing the state evolution of the system. By combining the idea of the heterogeneous mean-field theory with the effective degree approach, we are able to obtain the epidemic prevalence of the SIS process in uncorrelated static networks with arbitrary degree distributions with a higher precision than other approaches. It is worth noting that the epidemic threshold can be calculated as a corollary of the epidemic prevalence. Specifically, for SIS in scale-free networks, the quenched mean-field theory predicts that the epidemic threshold is zero in the thermodynamic limit. By contrast, our theoretical results show  that it remains  finite  for scale-free networks with $\gamma>3$, even in the thermodynamic limit.

Our work can be generalized to more general stochastic-logistic models of density dependent population dynamics~\cite{o2}. In such cases, both the infection and recovery rates usually depend on the ratio $I/N$ in these models (and $I$ is often interpreted as the population size, while $N$ is the carrying capacity). Our dynamic correlation approach could be fairly straightforwardly extended to the general stochastic logistic model where Eq.~(\ref{eq:mstr}) is the first equation to be modified (to account for the population-dependent death rate).

\bigskip
\textit{Note added.} When the main part of this paper is ready for publication in PRL, we received several useful comments from Dr. Silvio Ferreira. Specifically, Ferreira and coworkers have also considered the dynamic correlation problem of the SIS model (and also the contact process) in quenched static networks by introducing three-vertex approximation~\cite{Ferreira2013} and pair quenched mean-field theory~\cite{epl2013}. Particularly, by using a heterogeneous pair-approximation~\cite{NJP2014}, they also found that the epidemic threshold of the SIS process on static network reads as Eq.~(\ref{eq.9}). We thank Dr. Silvio Ferreira for bringing these interesting works to our attention. We would also like to point out that our theoretical analysis of the SIS model (which is mainly based on the detailed balance equilibrium condition and the maximal entropy principle) is devoted to explicitly derive analytical expression for the prevalence in the stationary state, and the epidemic threshold can then be calculated as a corollary of the prevalence. Based on some heuristic arguments, it was shown that small-world
random networks with a degree distribution decaying slower than an exponential have a vanishing epidemic threshold in the thermodynamic limit~\cite{PhysRevLett.111.068701}. This point was also set in a more general context in~\cite{Ferreira2016pre}. In addition, for the contact process in random graphs with power law degree distributions, a rigorous proof for the vanishing epidemic threshold was provided by Chatterjee and Durret in Ref.~\cite{Chatterjee2009}. From the result presented in Refs.~\cite{th,epl2013,PhysRevE.91.012816,PhysRevE.93.032322}, we observed that there are usually two peaks for the susceptibility against infection rate, which is obtained by quasistationary simulation of the SIS model in scale-free networks with $\gamma>3$. For this peculiar characteristic of the SIS process on scale-free networks, we argue that there might exist two epidemic thresholds (hinted by the two peaks): One (the left peak) corresponds to the case that the amount of infected nodes becoming from zero to nonzero in the long time limit (which also can be regarded as the localized state~\cite{PhysRevLett.109.128702}, i.e., the epidemic can only persist between connected hub nodes); The other (the right peak) corresponds to the case that the fraction of infected becomes from zero to nonzero in the thermodynamic limit (i.e., the fraction of infected could be kept as a non-zero stationary level). We think that, in localized state, the infected nodes are restricted to the hubs, and the fraction of infected becomes vanishing small in the thermodynamic limit. Thus, Our current understanding for the SIS model on (quenched) static networks from quasistationary simulations with increasing $\lambda$ could be summarized as follows: (1) For sufficiently small $\lambda$, there are null infected, i.e., the number of infected nodes goes to zero in the long time limit; (2) With the increase of $\lambda$, the epidemic can maintain active between hubs, and the nodes with small degrees are relatively difficult to be infected; (3) With the even increase of $\lambda$, the collective activation process emerges, i.e., the fraction of infected will maintain a nonzero-level in the thermodynamic limit. We thought that the studies implemented in Refs.~\cite{PhysRevLett.111.068701,Ferreira2016pre,Chatterjee2009} focus more on the process $(1) \rightarrow (2)$, while our proposed method and HMF theory focus more on the process $(2) \rightarrow (3)$. In this sense, we think that our current theoretical analysis gives reasonable prediction for the epidemic threshold (on scale free networks with $\gamma>3$), which suggests the system will transit from (2) to (3) with a finite threshold indicated by Eq.~(\ref{eq.9}). Once again, we would like to thank Dr. Ferreira for bringing these important literature to our attention and his instructive comments on the epidemic threshold of SIS model on heterogeneous networks.

\bigskip
\begin{acknowledgments}
This work was supported by the National Natural Science Foundation of China (Grants No.\ 11135001, No.\ 11575072, No.\ 11475074, and No.\ 61374053), and by the Fundamental Research Funds for the Central Universities (Grant No.\ lzujbky-2015-206). P.H. was supported by Basic Science Research Program through the National Research Foundation of Korea (NRF) funded by the Ministry of Education (2016R1D1A1B01007774).
\end{acknowledgments}
\bibliography{bbb}

\providecommand{\noopsort}[1]{}\providecommand{\singleletter}[1]{#1}%
\begin{thebibliography}{49}%
\makeatletter
\providecommand \@ifxundefined [1]{%
 \@ifx{#1\undefined}
}%
\providecommand \@ifnum [1]{%
 \ifnum #1\expandafter \@firstoftwo
 \else \expandafter \@secondoftwo
 \fi
}%
\providecommand \@ifx [1]{%
 \ifx #1\expandafter \@firstoftwo
 \else \expandafter \@secondoftwo
 \fi
}%
\providecommand \natexlab [1]{#1}%
\providecommand \enquote  [1]{``#1''}%
\providecommand \bibnamefont  [1]{#1}%
\providecommand \bibfnamefont [1]{#1}%
\providecommand \citenamefont [1]{#1}%
\providecommand \href@noop [0]{\@secondoftwo}%
\providecommand \href [0]{\begingroup \@sanitize@url \@href}%
\providecommand \@href[1]{\@@startlink{#1}\@@href}%
\providecommand \@@href[1]{\endgroup#1\@@endlink}%
\providecommand \@sanitize@url [0]{\catcode `\\12\catcode `\$12\catcode
  `\&12\catcode `\#12\catcode `\^12\catcode `\_12\catcode `\%12\relax}%
\providecommand \@@startlink[1]{}%
\providecommand \@@endlink[0]{}%
\providecommand \url  [0]{\begingroup\@sanitize@url \@url }%
\providecommand \@url [1]{\endgroup\@href {#1}{\urlprefix }}%
\providecommand \urlprefix  [0]{URL }%
\providecommand \Eprint [0]{\href }%
\providecommand \doibase [0]{http://dx.doi.org/}%
\providecommand \selectlanguage [0]{\@gobble}%
\providecommand \bibinfo  [0]{\@secondoftwo}%
\providecommand \bibfield  [0]{\@secondoftwo}%
\providecommand \translation [1]{[#1]}%
\providecommand \BibitemOpen [0]{}%
\providecommand \bibitemStop [0]{}%
\providecommand \bibitemNoStop [0]{.\EOS\space}%
\providecommand \EOS [0]{\spacefactor3000\relax}%
\providecommand \BibitemShut  [1]{\csname bibitem#1\endcsname}%
\let\auto@bib@innerbib\@empty
\bibitem [{\citenamefont {Keeling}\ and\ \citenamefont
  {Eames}(2005)}]{keeling_rev}%
  \BibitemOpen
  \bibfield  {author} {\bibinfo {author} {\bibfnamefont {M.~J.}\ \bibnamefont
  {Keeling}}\ and\ \bibinfo {author} {\bibfnamefont {K.~T.}\ \bibnamefont
  {Eames}},\ }\href {\doibase 10.1098/rsif.2005.0051} {\bibfield  {journal}
  {\bibinfo  {journal} {Journal of The Royal Society Interface}\ }\textbf
  {\bibinfo {volume} {2}},\ \bibinfo {pages} {295} (\bibinfo {year}
  {2005})}\BibitemShut {NoStop}%
\bibitem [{\citenamefont {Marro}\ and\ \citenamefont
  {Dickman}(1999)}]{Marrobook}%
  \BibitemOpen
  \bibfield  {author} {\bibinfo {author} {\bibfnamefont {J.}~\bibnamefont
  {Marro}}\ and\ \bibinfo {author} {\bibfnamefont {R.}~\bibnamefont
  {Dickman}},\ }\href@noop {} {\emph {\bibinfo {title} {Nonequilibrium Phase
  Transitions in Lattice Models}}}\ (\bibinfo  {publisher} {Cambridge
  University Press, Cambridge},\ \bibinfo {year} {1999})\BibitemShut {NoStop}%
\bibitem [{\citenamefont {Holme}(2015)}]{holme_sis}%
  \BibitemOpen
  \bibfield  {author} {\bibinfo {author} {\bibfnamefont {P.}~\bibnamefont
  {Holme}},\ }\href {\doibase 10.1103/PhysRevE.92.012804} {\bibfield  {journal}
  {\bibinfo  {journal} {Phys. Rev. E}\ }\textbf {\bibinfo {volume} {92}},\
  \bibinfo {pages} {012804} (\bibinfo {year} {2015})}\BibitemShut {NoStop}%
\bibitem [{\citenamefont {N{\aa}sell}(2001)}]{nasell}%
  \BibitemOpen
  \bibfield  {author} {\bibinfo {author} {\bibfnamefont {I.}~\bibnamefont
  {N{\aa}sell}},\ }\href {\doibase http://dx.doi.org/10.1006/jtbi.2001.2328}
  {\bibfield  {journal} {\bibinfo  {journal} {J. Theor. Biol.}\ }\textbf
  {\bibinfo {volume} {211}},\ \bibinfo {pages} {11 } (\bibinfo {year}
  {2001})}\BibitemShut {NoStop}%
\bibitem [{\citenamefont {Pastor-Satorras}\ and\ \citenamefont
  {Vespignani}(2001)}]{hmf_1}%
  \BibitemOpen
  \bibfield  {author} {\bibinfo {author} {\bibfnamefont {R.}~\bibnamefont
  {Pastor-Satorras}}\ and\ \bibinfo {author} {\bibfnamefont {A.}~\bibnamefont
  {Vespignani}},\ }\href {\doibase 10.1103/PhysRevLett.86.3200} {\bibfield
  {journal} {\bibinfo  {journal} {Phys. Rev. Lett.}\ }\textbf {\bibinfo
  {volume} {86}},\ \bibinfo {pages} {3200} (\bibinfo {year}
  {2001})}\BibitemShut {NoStop}%
\bibitem [{\citenamefont {Barrat}\ \emph {et~al.}(2008)\citenamefont {Barrat},
  \citenamefont {Barth\'elemy},\ and\ \citenamefont {Vespignani}}]{hmf_3}%
  \BibitemOpen
  \bibfield  {author} {\bibinfo {author} {\bibfnamefont {A.}~\bibnamefont
  {Barrat}}, \bibinfo {author} {\bibfnamefont {M.}~\bibnamefont
  {Barth\'elemy}}, \ and\ \bibinfo {author} {\bibfnamefont {A.}~\bibnamefont
  {Vespignani}},\ }\href@noop {} {\emph {\bibinfo {title} {Dynamical Processes
  on Complex Networks}}}\ (\bibinfo  {publisher} {Cambridge University Press,
  Cambridge},\ \bibinfo {year} {2008})\BibitemShut {NoStop}%
\bibitem [{\citenamefont {Hethcote}(2000)}]{hethcoterev}%
  \BibitemOpen
  \bibfield  {author} {\bibinfo {author} {\bibfnamefont {H.~W.}\ \bibnamefont
  {Hethcote}},\ }\href {\doibase 10.1137/S0036144500371907} {\bibfield
  {journal} {\bibinfo  {journal} {SIAM Rev.}\ }\textbf {\bibinfo {volume}
  {42}},\ \bibinfo {pages} {599} (\bibinfo {year} {2000})}\BibitemShut
  {NoStop}%
\bibitem [{\citenamefont {Albert}\ and\ \citenamefont
  {Barab\'asi}(2002)}]{rmp2002}%
  \BibitemOpen
  \bibfield  {author} {\bibinfo {author} {\bibfnamefont {R.}~\bibnamefont
  {Albert}}\ and\ \bibinfo {author} {\bibfnamefont {A.-L.}\ \bibnamefont
  {Barab\'asi}},\ }\href {\doibase 10.1103/RevModPhys.74.47} {\bibfield
  {journal} {\bibinfo  {journal} {Rev. Mod. Phys.}\ }\textbf {\bibinfo {volume}
  {74}},\ \bibinfo {pages} {47} (\bibinfo {year} {2002})}\BibitemShut {NoStop}%
\bibitem [{\citenamefont {Dorogovtsev}\ \emph {et~al.}(2008)\citenamefont
  {Dorogovtsev}, \citenamefont {Goltsev},\ and\ \citenamefont
  {Mendes}}]{hmf_2}%
  \BibitemOpen
  \bibfield  {author} {\bibinfo {author} {\bibfnamefont {S.~N.}\ \bibnamefont
  {Dorogovtsev}}, \bibinfo {author} {\bibfnamefont {A.~V.}\ \bibnamefont
  {Goltsev}}, \ and\ \bibinfo {author} {\bibfnamefont {J.~F.~F.}\ \bibnamefont
  {Mendes}},\ }\href {\doibase 10.1103/RevModPhys.80.1275} {\bibfield
  {journal} {\bibinfo  {journal} {Rev. Mod. Phys.}\ }\textbf {\bibinfo {volume}
  {80}},\ \bibinfo {pages} {1275} (\bibinfo {year} {2008})}\BibitemShut
  {NoStop}%
\bibitem [{\citenamefont {Ferreira}\ \emph {et~al.}(2012)\citenamefont
  {Ferreira}, \citenamefont {Castellano},\ and\ \citenamefont
  {Pastor-Satorras}}]{th}%
  \BibitemOpen
  \bibfield  {author} {\bibinfo {author} {\bibfnamefont {S.~C.}\ \bibnamefont
  {Ferreira}}, \bibinfo {author} {\bibfnamefont {C.}~\bibnamefont
  {Castellano}}, \ and\ \bibinfo {author} {\bibfnamefont {R.}~\bibnamefont
  {Pastor-Satorras}},\ }\href {\doibase 10.1103/PhysRevE.86.041125} {\bibfield
  {journal} {\bibinfo  {journal} {Phys. Rev. E}\ }\textbf {\bibinfo {volume}
  {86}},\ \bibinfo {pages} {041125} (\bibinfo {year} {2012})}\BibitemShut
  {NoStop}%
\bibitem [{\citenamefont {Pastor-Satorras}\ \emph {et~al.}(2015)\citenamefont
  {Pastor-Satorras}, \citenamefont {Castellano}, \citenamefont {Van~Mieghem},\
  and\ \citenamefont {Vespignani}}]{ss_2}%
  \BibitemOpen
  \bibfield  {author} {\bibinfo {author} {\bibfnamefont {R.}~\bibnamefont
  {Pastor-Satorras}}, \bibinfo {author} {\bibfnamefont {C.}~\bibnamefont
  {Castellano}}, \bibinfo {author} {\bibfnamefont {P.}~\bibnamefont
  {Van~Mieghem}}, \ and\ \bibinfo {author} {\bibfnamefont {A.}~\bibnamefont
  {Vespignani}},\ }\href {\doibase 10.1103/RevModPhys.87.925} {\bibfield
  {journal} {\bibinfo  {journal} {Rev. Mod. Phys.}\ }\textbf {\bibinfo {volume}
  {87}},\ \bibinfo {pages} {925} (\bibinfo {year} {2015})}\BibitemShut
  {NoStop}%
\bibitem [{sm()}]{sm}%
  \BibitemOpen
  \href@noop {} {}\bibinfo {note} {See Supplemental Material [url] for brief
  reviews on the popular analytical treatments of the SIR and SIS processes on
  networks, and detailed algorithms and more results of our model, which
  includes
  Refs.~\cite{Callaway2000prl,Cohen2000prl,Oliveira2005pre,UCM}}\BibitemShut
  {NoStop}%
\bibitem [{\citenamefont {Van~Mieghem}\ \emph {et~al.}(2009)\citenamefont
  {Van~Mieghem}, \citenamefont {Omic},\ and\ \citenamefont {Kooij}}]{qmf_1}%
  \BibitemOpen
  \bibfield  {author} {\bibinfo {author} {\bibfnamefont {P.}~\bibnamefont
  {Van~Mieghem}}, \bibinfo {author} {\bibfnamefont {J.}~\bibnamefont {Omic}}, \
  and\ \bibinfo {author} {\bibfnamefont {R.}~\bibnamefont {Kooij}},\ }\href
  {\doibase 10.1109/TNET.2008.925623} {\bibfield  {journal} {\bibinfo
  {journal} {IEEE ACM Trans. Netw.}\ }\textbf {\bibinfo {volume} {17}},\
  \bibinfo {pages} {1} (\bibinfo {year} {2009})}\BibitemShut {NoStop}%
\bibitem [{\citenamefont {Granell}\ \emph {et~al.}(2013)\citenamefont
  {Granell}, \citenamefont {G\'omez},\ and\ \citenamefont {Arenas}}]{qmf_2}%
  \BibitemOpen
  \bibfield  {author} {\bibinfo {author} {\bibfnamefont {C.}~\bibnamefont
  {Granell}}, \bibinfo {author} {\bibfnamefont {S.}~\bibnamefont {G\'omez}}, \
  and\ \bibinfo {author} {\bibfnamefont {A.}~\bibnamefont {Arenas}},\ }\href
  {\doibase 10.1103/PhysRevLett.111.128701} {\bibfield  {journal} {\bibinfo
  {journal} {Phys. Rev. Lett.}\ }\textbf {\bibinfo {volume} {111}},\ \bibinfo
  {pages} {128701} (\bibinfo {year} {2013})}\BibitemShut {NoStop}%
\bibitem [{\citenamefont {Chung}\ \emph {et~al.}(2003)\citenamefont {Chung},
  \citenamefont {Lu},\ and\ \citenamefont {Vu}}]{Chung27052003}%
  \BibitemOpen
  \bibfield  {author} {\bibinfo {author} {\bibfnamefont {F.}~\bibnamefont
  {Chung}}, \bibinfo {author} {\bibfnamefont {L.}~\bibnamefont {Lu}}, \ and\
  \bibinfo {author} {\bibfnamefont {V.}~\bibnamefont {Vu}},\ }\href {\doibase
  10.1073/pnas.0937490100} {\bibfield  {journal} {\bibinfo  {journal} {Proc.
  Natl. Acad. Sci. U.S.A.}\ }\textbf {\bibinfo {volume} {100}},\ \bibinfo
  {pages} {6313} (\bibinfo {year} {2003})}\BibitemShut {NoStop}%
\bibitem [{\citenamefont {Castellano}\ and\ \citenamefont
  {Pastor-Satorras}(2010)}]{Claudio2010prl}%
  \BibitemOpen
  \bibfield  {author} {\bibinfo {author} {\bibfnamefont {C.}~\bibnamefont
  {Castellano}}\ and\ \bibinfo {author} {\bibfnamefont {R.}~\bibnamefont
  {Pastor-Satorras}},\ }\href {\doibase 10.1103/PhysRevLett.105.218701}
  {\bibfield  {journal} {\bibinfo  {journal} {Phys. Rev. Lett.}\ }\textbf
  {\bibinfo {volume} {105}},\ \bibinfo {pages} {218701} (\bibinfo {year}
  {2010})}\BibitemShut {NoStop}%
\bibitem [{\citenamefont {Bogu\~n\'a}\ \emph {et~al.}(2013)\citenamefont
  {Bogu\~n\'a}, \citenamefont {Castellano},\ and\ \citenamefont
  {Pastor-Satorras}}]{PhysRevLett.111.068701}%
  \BibitemOpen
  \bibfield  {author} {\bibinfo {author} {\bibfnamefont {M.}~\bibnamefont
  {Bogu\~n\'a}}, \bibinfo {author} {\bibfnamefont {C.}~\bibnamefont
  {Castellano}}, \ and\ \bibinfo {author} {\bibfnamefont {R.}~\bibnamefont
  {Pastor-Satorras}},\ }\href {\doibase 10.1103/PhysRevLett.111.068701}
  {\bibfield  {journal} {\bibinfo  {journal} {Phys. Rev. Lett.}\ }\textbf
  {\bibinfo {volume} {111}},\ \bibinfo {pages} {068701} (\bibinfo {year}
  {2013})}\BibitemShut {NoStop}%
\bibitem [{\citenamefont {Lindquist}\ \emph {et~al.}(2011)\citenamefont
  {Lindquist}, \citenamefont {Ma}, \citenamefont {Driessche},\ and\
  \citenamefont {Willeboordse}}]{ed_1}%
  \BibitemOpen
  \bibfield  {author} {\bibinfo {author} {\bibfnamefont {J.}~\bibnamefont
  {Lindquist}}, \bibinfo {author} {\bibfnamefont {J.}~\bibnamefont {Ma}},
  \bibinfo {author} {\bibfnamefont {P.}~\bibnamefont {Driessche}}, \ and\
  \bibinfo {author} {\bibfnamefont {F.}~\bibnamefont {Willeboordse}},\ }\href
  {\doibase 10.1007/s00285-010-0331-2} {\bibfield  {journal} {\bibinfo
  {journal} {J. Math. Biol.}\ }\textbf {\bibinfo {volume} {62}},\ \bibinfo
  {pages} {143} (\bibinfo {year} {2011})}\BibitemShut {NoStop}%
\bibitem [{\citenamefont {Cai}\ \emph {et~al.}(2014)\citenamefont {Cai},
  \citenamefont {Wu},\ and\ \citenamefont {Guan}}]{ed_2}%
  \BibitemOpen
  \bibfield  {author} {\bibinfo {author} {\bibfnamefont {C.-R.}\ \bibnamefont
  {Cai}}, \bibinfo {author} {\bibfnamefont {Z.-X.}\ \bibnamefont {Wu}}, \ and\
  \bibinfo {author} {\bibfnamefont {J.-Y.}\ \bibnamefont {Guan}},\ }\href
  {\doibase 10.1103/PhysRevE.90.052803} {\bibfield  {journal} {\bibinfo
  {journal} {Phys. Rev. E}\ }\textbf {\bibinfo {volume} {90}},\ \bibinfo
  {pages} {052803} (\bibinfo {year} {2014})}\BibitemShut {NoStop}%
\bibitem [{\citenamefont {Grassberger}(1983)}]{Grassberger}%
  \BibitemOpen
  \bibfield  {author} {\bibinfo {author} {\bibfnamefont {P.}~\bibnamefont
  {Grassberger}},\ }\href {\doibase
  http://dx.doi.org/10.1016/0025-5564(82)90036-0} {\bibfield  {journal}
  {\bibinfo  {journal} {Math. Biosci.}\ }\textbf {\bibinfo {volume} {63}},\
  \bibinfo {pages} {157 } (\bibinfo {year} {1983})}\BibitemShut {NoStop}%
\bibitem [{\citenamefont {Goltsev}\ \emph {et~al.}(2012)\citenamefont
  {Goltsev}, \citenamefont {Dorogovtsev}, \citenamefont {Oliveira},\ and\
  \citenamefont {Mendes}}]{PhysRevLett.109.128702}%
  \BibitemOpen
  \bibfield  {author} {\bibinfo {author} {\bibfnamefont {A.~V.}\ \bibnamefont
  {Goltsev}}, \bibinfo {author} {\bibfnamefont {S.~N.}\ \bibnamefont
  {Dorogovtsev}}, \bibinfo {author} {\bibfnamefont {J.~G.}\ \bibnamefont
  {Oliveira}}, \ and\ \bibinfo {author} {\bibfnamefont {J.~F.~F.}\ \bibnamefont
  {Mendes}},\ }\href {\doibase 10.1103/PhysRevLett.109.128702} {\bibfield
  {journal} {\bibinfo  {journal} {Phys. Rev. Lett.}\ }\textbf {\bibinfo
  {volume} {109}},\ \bibinfo {pages} {128702} (\bibinfo {year}
  {2012})}\BibitemShut {NoStop}%
\bibitem [{\citenamefont {Ferreira}\ \emph {et~al.}(2016)\citenamefont
  {Ferreira}, \citenamefont {Sander},\ and\ \citenamefont
  {Pastor-Satorras}}]{Ferreira2016pre}%
  \BibitemOpen
  \bibfield  {author} {\bibinfo {author} {\bibfnamefont {S.~C.}\ \bibnamefont
  {Ferreira}}, \bibinfo {author} {\bibfnamefont {R.~S.}\ \bibnamefont
  {Sander}}, \ and\ \bibinfo {author} {\bibfnamefont {R.}~\bibnamefont
  {Pastor-Satorras}},\ }\href {\doibase 10.1103/PhysRevE.93.032314} {\bibfield
  {journal} {\bibinfo  {journal} {Phys. Rev. E}\ }\textbf {\bibinfo {volume}
  {93}},\ \bibinfo {pages} {032314} (\bibinfo {year} {2016})}\BibitemShut
  {NoStop}%
\bibitem [{\citenamefont {Ruhi}\ \emph {et~al.}()\citenamefont {Ruhi},
  \citenamefont {Thrampoulidis},\ and\ \citenamefont {Hassibi}}]{ruhi}%
  \BibitemOpen
  \bibfield  {author} {\bibinfo {author} {\bibfnamefont {N.~A.}\ \bibnamefont
  {Ruhi}}, \bibinfo {author} {\bibfnamefont {C.}~\bibnamefont {Thrampoulidis}},
  \ and\ \bibinfo {author} {\bibfnamefont {B.}~\bibnamefont {Hassibi}},\ }\href
  {http://arxiv.org/abs/1603.05095} {}\bibinfo {note} {E-print
  arXiv:1603.05095}\BibitemShut {NoStop}%
\bibitem [{\citenamefont {Lee}\ \emph {et~al.}(2013)\citenamefont {Lee},
  \citenamefont {Shim},\ and\ \citenamefont {Noh}}]{PhysRevE.87.062812}%
  \BibitemOpen
  \bibfield  {author} {\bibinfo {author} {\bibfnamefont {H.~K.}\ \bibnamefont
  {Lee}}, \bibinfo {author} {\bibfnamefont {P.-S.}\ \bibnamefont {Shim}}, \
  and\ \bibinfo {author} {\bibfnamefont {J.~D.}\ \bibnamefont {Noh}},\ }\href
  {\doibase 10.1103/PhysRevE.87.062812} {\bibfield  {journal} {\bibinfo
  {journal} {Phys. Rev. E}\ }\textbf {\bibinfo {volume} {87}},\ \bibinfo
  {pages} {062812} (\bibinfo {year} {2013})}\BibitemShut {NoStop}%
\bibitem [{\citenamefont {Eames}\ and\ \citenamefont {Keeling}(2002)}]{pa_1}%
  \BibitemOpen
  \bibfield  {author} {\bibinfo {author} {\bibfnamefont {K.~T.~D.}\
  \bibnamefont {Eames}}\ and\ \bibinfo {author} {\bibfnamefont {M.~J.}\
  \bibnamefont {Keeling}},\ }\href {\doibase 10.1073/pnas.202244299} {\bibfield
   {journal} {\bibinfo  {journal} {Proc. Natl. Acad. Sci. U.S.A.}\ }\textbf
  {\bibinfo {volume} {99}},\ \bibinfo {pages} {13330} (\bibinfo {year}
  {2002})}\BibitemShut {NoStop}%
\bibitem [{\citenamefont {Kiss}\ \emph {et~al.}(2015)\citenamefont {Kiss},
  \citenamefont {R\"ost},\ and\ \citenamefont {Vizi}}]{pa_2}%
  \BibitemOpen
  \bibfield  {author} {\bibinfo {author} {\bibfnamefont {I.~Z.}\ \bibnamefont
  {Kiss}}, \bibinfo {author} {\bibfnamefont {G.}~\bibnamefont {R\"ost}}, \ and\
  \bibinfo {author} {\bibfnamefont {Z.}~\bibnamefont {Vizi}},\ }\href {\doibase
  10.1103/PhysRevLett.115.078701} {\bibfield  {journal} {\bibinfo  {journal}
  {Phys. Rev. Lett.}\ }\textbf {\bibinfo {volume} {115}},\ \bibinfo {pages}
  {078701} (\bibinfo {year} {2015})}\BibitemShut {NoStop}%
\bibitem [{\citenamefont {Newman}(2002)}]{sn_1}%
  \BibitemOpen
  \bibfield  {author} {\bibinfo {author} {\bibfnamefont {M.~E.~J.}\
  \bibnamefont {Newman}},\ }\href {\doibase 10.1103/PhysRevE.66.016128}
  {\bibfield  {journal} {\bibinfo  {journal} {Phys. Rev. E}\ }\textbf {\bibinfo
  {volume} {66}},\ \bibinfo {pages} {016128} (\bibinfo {year}
  {2002})}\BibitemShut {NoStop}%
\bibitem [{\citenamefont {Volz}\ and\ \citenamefont {Meyers}(2007)}]{sn_2}%
  \BibitemOpen
  \bibfield  {author} {\bibinfo {author} {\bibfnamefont {E.}~\bibnamefont
  {Volz}}\ and\ \bibinfo {author} {\bibfnamefont {L.~A.}\ \bibnamefont
  {Meyers}},\ }\href {\doibase 10.1098/rspb.2007.1159} {\bibfield  {journal}
  {\bibinfo  {journal} {Pro. R. Soc. London Ser. B}\ }\textbf {\bibinfo
  {volume} {274}},\ \bibinfo {pages} {2925} (\bibinfo {year}
  {2007})}\BibitemShut {NoStop}%
\bibitem [{\citenamefont {Aparicio}\ and\ \citenamefont
  {Pascual}(2007)}]{sn_3}%
  \BibitemOpen
  \bibfield  {author} {\bibinfo {author} {\bibfnamefont {J.~P.}\ \bibnamefont
  {Aparicio}}\ and\ \bibinfo {author} {\bibfnamefont {M.}~\bibnamefont
  {Pascual}},\ }\href {\doibase 10.1098/rspb.2006.0057} {\bibfield  {journal}
  {\bibinfo  {journal} {Pro. R. Soc. London Ser. B}\ }\textbf {\bibinfo
  {volume} {274}},\ \bibinfo {pages} {505} (\bibinfo {year}
  {2007})}\BibitemShut {NoStop}%
\bibitem [{\citenamefont {Parshani}\ \emph {et~al.}(2010)\citenamefont
  {Parshani}, \citenamefont {Carmi},\ and\ \citenamefont
  {Havlin}}]{PhysRevLett.104.258701}%
  \BibitemOpen
  \bibfield  {author} {\bibinfo {author} {\bibfnamefont {R.}~\bibnamefont
  {Parshani}}, \bibinfo {author} {\bibfnamefont {S.}~\bibnamefont {Carmi}}, \
  and\ \bibinfo {author} {\bibfnamefont {S.}~\bibnamefont {Havlin}},\ }\href
  {\doibase 10.1103/PhysRevLett.104.258701} {\bibfield  {journal} {\bibinfo
  {journal} {Phys. Rev. Lett.}\ }\textbf {\bibinfo {volume} {104}},\ \bibinfo
  {pages} {258701} (\bibinfo {year} {2010})}\BibitemShut {NoStop}%
\bibitem [{\citenamefont {Leventhal}\ \emph {et~al.}(2015)\citenamefont
  {Leventhal}, \citenamefont {Hill}, \citenamefont {Nowak},\ and\ \citenamefont
  {Bonhoeffer}}]{NC}%
  \BibitemOpen
  \bibfield  {author} {\bibinfo {author} {\bibfnamefont {G.~E.}\ \bibnamefont
  {Leventhal}}, \bibinfo {author} {\bibfnamefont {A.~L.}\ \bibnamefont {Hill}},
  \bibinfo {author} {\bibfnamefont {M.~A.}\ \bibnamefont {Nowak}}, \ and\
  \bibinfo {author} {\bibfnamefont {S.}~\bibnamefont {Bonhoeffer}},\ }\href
  {\doibase 10.1038/ncomms7101} {\bibfield  {journal} {\bibinfo  {journal}
  {Nat. Commun.}\ }\textbf {\bibinfo {volume} {6}},\ \bibinfo {pages} {6101}
  (\bibinfo {year} {2015})}\BibitemShut {NoStop}%
\bibitem [{\citenamefont {Gleeson}(2011)}]{Gleeson_1}%
  \BibitemOpen
  \bibfield  {author} {\bibinfo {author} {\bibfnamefont {J.~P.}\ \bibnamefont
  {Gleeson}},\ }\href {\doibase 10.1103/PhysRevLett.107.068701} {\bibfield
  {journal} {\bibinfo  {journal} {Phys. Rev. Lett.}\ }\textbf {\bibinfo
  {volume} {107}},\ \bibinfo {pages} {068701} (\bibinfo {year}
  {2011})}\BibitemShut {NoStop}%
\bibitem [{\citenamefont {Gleeson}(2013)}]{Gleeson_2}%
  \BibitemOpen
  \bibfield  {author} {\bibinfo {author} {\bibfnamefont {J.~P.}\ \bibnamefont
  {Gleeson}},\ }\href {\doibase 10.1103/PhysRevX.3.021004} {\bibfield
  {journal} {\bibinfo  {journal} {Phys. Rev. X}\ }\textbf {\bibinfo {volume}
  {3}},\ \bibinfo {pages} {021004} (\bibinfo {year} {2013})}\BibitemShut
  {NoStop}%
\bibitem [{\citenamefont {Keeling}(1999)}]{keeling}%
  \BibitemOpen
  \bibfield  {author} {\bibinfo {author} {\bibfnamefont {M.~J.}\ \bibnamefont
  {Keeling}},\ }\href {\doibase 10.1098/rspb.1999.0716} {\bibfield  {journal}
  {\bibinfo  {journal} {Pro. R. Soc. London Ser. B}\ }\textbf {\bibinfo
  {volume} {266}},\ \bibinfo {pages} {859} (\bibinfo {year}
  {1999})}\BibitemShut {NoStop}%
\bibitem [{\citenamefont {Jaynes}(1957)}]{Jaynes1957}%
  \BibitemOpen
  \bibfield  {author} {\bibinfo {author} {\bibfnamefont {E.~T.}\ \bibnamefont
  {Jaynes}},\ }\href {\doibase 10.1103/PhysRev.106.620} {\bibfield  {journal}
  {\bibinfo  {journal} {Phys. Rev.}\ }\textbf {\bibinfo {volume} {106}},\
  \bibinfo {pages} {620} (\bibinfo {year} {1957})}\BibitemShut {NoStop}%
\bibitem [{\citenamefont {Castellano}\ and\ \citenamefont
  {Pastor-Satorras}(2006)}]{PhysRevLett.96.038701}%
  \BibitemOpen
  \bibfield  {author} {\bibinfo {author} {\bibfnamefont {C.}~\bibnamefont
  {Castellano}}\ and\ \bibinfo {author} {\bibfnamefont {R.}~\bibnamefont
  {Pastor-Satorras}},\ }\href {\doibase 10.1103/PhysRevLett.96.038701}
  {\bibfield  {journal} {\bibinfo  {journal} {Phys. Rev. Lett.}\ }\textbf
  {\bibinfo {volume} {96}},\ \bibinfo {pages} {038701} (\bibinfo {year}
  {2006})}\BibitemShut {NoStop}%
\bibitem [{\citenamefont {Ferreira}\ \emph {et~al.}(2011)\citenamefont
  {Ferreira}, \citenamefont {Ferreira}, \citenamefont {Castellano},\ and\
  \citenamefont {Pastor-Satorras}}]{PhysRevE.84.066102}%
  \BibitemOpen
  \bibfield  {author} {\bibinfo {author} {\bibfnamefont {S.~C.}\ \bibnamefont
  {Ferreira}}, \bibinfo {author} {\bibfnamefont {R.~S.}\ \bibnamefont
  {Ferreira}}, \bibinfo {author} {\bibfnamefont {C.}~\bibnamefont
  {Castellano}}, \ and\ \bibinfo {author} {\bibfnamefont {R.}~\bibnamefont
  {Pastor-Satorras}},\ }\href {\doibase 10.1103/PhysRevE.84.066102} {\bibfield
  {journal} {\bibinfo  {journal} {Phys. Rev. E}\ }\textbf {\bibinfo {volume}
  {84}},\ \bibinfo {pages} {066102} (\bibinfo {year} {2011})}\BibitemShut
  {NoStop}%
\bibitem [{\citenamefont {Mu\~noz}\ \emph {et~al.}(2010)\citenamefont
  {Mu\~noz}, \citenamefont {Juh\'asz}, \citenamefont {Castellano},\ and\
  \citenamefont {\'Odor}}]{cp2010prl}%
  \BibitemOpen
  \bibfield  {author} {\bibinfo {author} {\bibfnamefont {M.~A.}\ \bibnamefont
  {Mu\~noz}}, \bibinfo {author} {\bibfnamefont {R.}~\bibnamefont {Juh\'asz}},
  \bibinfo {author} {\bibfnamefont {C.}~\bibnamefont {Castellano}}, \ and\
  \bibinfo {author} {\bibfnamefont {G.}~\bibnamefont {\'Odor}},\ }\href
  {\doibase 10.1103/PhysRevLett.105.128701} {\bibfield  {journal} {\bibinfo
  {journal} {Phys. Rev. Lett.}\ }\textbf {\bibinfo {volume} {105}},\ \bibinfo
  {pages} {128701} (\bibinfo {year} {2010})}\BibitemShut {NoStop}%
\bibitem [{\citenamefont {Ovaskainen}\ and\ \citenamefont
  {Meerson}(2010)}]{o2}%
  \BibitemOpen
  \bibfield  {author} {\bibinfo {author} {\bibfnamefont {O.}~\bibnamefont
  {Ovaskainen}}\ and\ \bibinfo {author} {\bibfnamefont {B.}~\bibnamefont
  {Meerson}},\ }\href {\doibase http://dx.doi.org/10.1016/j.tree.2010.07.009}
  {\bibfield  {journal} {\bibinfo  {journal} {Trend. Ecol. Evol.}\ }\textbf
  {\bibinfo {volume} {25}},\ \bibinfo {pages} {643 } (\bibinfo {year}
  {2010})}\BibitemShut {NoStop}%
\bibitem [{\citenamefont {Ferreira}\ and\ \citenamefont
  {Ferreira}(2013)}]{Ferreira2013}%
  \BibitemOpen
  \bibfield  {author} {\bibinfo {author} {\bibfnamefont {R.~S.}\ \bibnamefont
  {Ferreira}}\ and\ \bibinfo {author} {\bibfnamefont {S.~C.}\ \bibnamefont
  {Ferreira}},\ }\href {\doibase 10.1140/epjb/e2013-40534-0} {\bibfield
  {journal} {\bibinfo  {journal} {The European Physical Journal B}\ }\textbf
  {\bibinfo {volume} {86}},\ \bibinfo {pages} {1} (\bibinfo {year}
  {2013})}\BibitemShut {NoStop}%
\bibitem [{\citenamefont {{Mata, Ang¨¦lica S.}}\ and\ \citenamefont {{Ferreira,
  Silvio C.}}(2013)}]{epl2013}%
  \BibitemOpen
  \bibfield  {author} {\bibinfo {author} {\bibnamefont {{Mata, Ang¨¦lica S.}}}\
  and\ \bibinfo {author} {\bibnamefont {{Ferreira, Silvio C.}}},\ }\href
  {\doibase 10.1209/0295-5075/103/48003} {\bibfield  {journal} {\bibinfo
  {journal} {EPL}\ }\textbf {\bibinfo {volume} {103}},\ \bibinfo {pages}
  {48003} (\bibinfo {year} {2013})}\BibitemShut {NoStop}%
\bibitem [{\citenamefont {Mata}\ \emph {et~al.}(2014)\citenamefont {Mata},
  \citenamefont {Ferreira},\ and\ \citenamefont {Ferreira}}]{NJP2014}%
  \BibitemOpen
  \bibfield  {author} {\bibinfo {author} {\bibfnamefont {A.~S.}\ \bibnamefont
  {Mata}}, \bibinfo {author} {\bibfnamefont {R.~S.}\ \bibnamefont {Ferreira}},
  \ and\ \bibinfo {author} {\bibfnamefont {S.~C.}\ \bibnamefont {Ferreira}},\
  }\href {http://stacks.iop.org/1367-2630/16/i=5/a=053006} {\bibfield
  {journal} {\bibinfo  {journal} {New Journal of Physics}\ }\textbf {\bibinfo
  {volume} {16}},\ \bibinfo {pages} {053006} (\bibinfo {year}
  {2014})}\BibitemShut {NoStop}%
\bibitem [{\citenamefont {Chatterjee}\ and\ \citenamefont
  {Durrett}(2009)}]{Chatterjee2009}%
  \BibitemOpen
  \bibfield  {author} {\bibinfo {author} {\bibfnamefont {S.}~\bibnamefont
  {Chatterjee}}\ and\ \bibinfo {author} {\bibfnamefont {R.}~\bibnamefont
  {Durrett}},\ }\href {\doibase 10.1214/09-AOP471} {\bibfield  {journal}
  {\bibinfo  {journal} {Ann. Probab.}\ }\textbf {\bibinfo {volume} {37}},\
  \bibinfo {pages} {2332} (\bibinfo {year} {2009})}\BibitemShut {NoStop}%
\bibitem [{\citenamefont {Mata}\ and\ \citenamefont
  {Ferreira}(2015)}]{PhysRevE.91.012816}%
  \BibitemOpen
  \bibfield  {author} {\bibinfo {author} {\bibfnamefont {A.~S.}\ \bibnamefont
  {Mata}}\ and\ \bibinfo {author} {\bibfnamefont {S.~C.}\ \bibnamefont
  {Ferreira}},\ }\href {\doibase 10.1103/PhysRevE.91.012816} {\bibfield
  {journal} {\bibinfo  {journal} {Phys. Rev. E}\ }\textbf {\bibinfo {volume}
  {91}},\ \bibinfo {pages} {012816} (\bibinfo {year} {2015})}\BibitemShut
  {NoStop}%
\bibitem [{\citenamefont {Cota}\ \emph {et~al.}(2016)\citenamefont {Cota},
  \citenamefont {Ferreira},\ and\ \citenamefont {\'Odor}}]{PhysRevE.93.032322}%
  \BibitemOpen
  \bibfield  {author} {\bibinfo {author} {\bibfnamefont {W.}~\bibnamefont
  {Cota}}, \bibinfo {author} {\bibfnamefont {S.~C.}\ \bibnamefont {Ferreira}},
  \ and\ \bibinfo {author} {\bibfnamefont {G.}~\bibnamefont {\'Odor}},\ }\href
  {\doibase 10.1103/PhysRevE.93.032322} {\bibfield  {journal} {\bibinfo
  {journal} {Phys. Rev. E}\ }\textbf {\bibinfo {volume} {93}},\ \bibinfo
  {pages} {032322} (\bibinfo {year} {2016})}\BibitemShut {NoStop}%
\bibitem [{\citenamefont {Callaway}\ \emph {et~al.}(2000)\citenamefont
  {Callaway}, \citenamefont {Newman}, \citenamefont {Strogatz},\ and\
  \citenamefont {Watts}}]{Callaway2000prl}%
  \BibitemOpen
  \bibfield  {author} {\bibinfo {author} {\bibfnamefont {D.~S.}\ \bibnamefont
  {Callaway}}, \bibinfo {author} {\bibfnamefont {M.~E.~J.}\ \bibnamefont
  {Newman}}, \bibinfo {author} {\bibfnamefont {S.~H.}\ \bibnamefont
  {Strogatz}}, \ and\ \bibinfo {author} {\bibfnamefont {D.~J.}\ \bibnamefont
  {Watts}},\ }\href {\doibase 10.1103/PhysRevLett.85.5468} {\bibfield
  {journal} {\bibinfo  {journal} {Phys. Rev. Lett.}\ }\textbf {\bibinfo
  {volume} {85}},\ \bibinfo {pages} {5468} (\bibinfo {year}
  {2000})}\BibitemShut {NoStop}%
\bibitem [{\citenamefont {Cohen}\ \emph {et~al.}(2000)\citenamefont {Cohen},
  \citenamefont {Erez}, \citenamefont {ben Avraham},\ and\ \citenamefont
  {Havlin}}]{Cohen2000prl}%
  \BibitemOpen
  \bibfield  {author} {\bibinfo {author} {\bibfnamefont {R.}~\bibnamefont
  {Cohen}}, \bibinfo {author} {\bibfnamefont {K.}~\bibnamefont {Erez}},
  \bibinfo {author} {\bibfnamefont {D.}~\bibnamefont {ben Avraham}}, \ and\
  \bibinfo {author} {\bibfnamefont {S.}~\bibnamefont {Havlin}},\ }\href
  {\doibase 10.1103/PhysRevLett.85.4626} {\bibfield  {journal} {\bibinfo
  {journal} {Phys. Rev. Lett.}\ }\textbf {\bibinfo {volume} {85}},\ \bibinfo
  {pages} {4626} (\bibinfo {year} {2000})}\BibitemShut {NoStop}%
\bibitem [{\citenamefont {de~Oliveira}\ and\ \citenamefont
  {Dickman}(2005)}]{Oliveira2005pre}%
  \BibitemOpen
  \bibfield  {author} {\bibinfo {author} {\bibfnamefont {M.~M.}\ \bibnamefont
  {de~Oliveira}}\ and\ \bibinfo {author} {\bibfnamefont {R.}~\bibnamefont
  {Dickman}},\ }\href {\doibase 10.1103/PhysRevE.71.016129} {\bibfield
  {journal} {\bibinfo  {journal} {Phys. Rev. E}\ }\textbf {\bibinfo {volume}
  {71}},\ \bibinfo {pages} {016129} (\bibinfo {year} {2005})}\BibitemShut
  {NoStop}%
\bibitem [{\citenamefont {Catanzaro}\ \emph {et~al.}(2005)\citenamefont
  {Catanzaro}, \citenamefont {Bogu\~n\'a},\ and\ \citenamefont
  {Pastor-Satorras}}]{UCM}%
  \BibitemOpen
  \bibfield  {author} {\bibinfo {author} {\bibfnamefont {M.}~\bibnamefont
  {Catanzaro}}, \bibinfo {author} {\bibfnamefont {M.}~\bibnamefont
  {Bogu\~n\'a}}, \ and\ \bibinfo {author} {\bibfnamefont {R.}~\bibnamefont
  {Pastor-Satorras}},\ }\href {\doibase 10.1103/PhysRevE.71.027103} {\bibfield
  {journal} {\bibinfo  {journal} {Phys. Rev. E}\ }\textbf {\bibinfo {volume}
  {71}},\ \bibinfo {pages} {027103} (\bibinfo {year} {2005})}\BibitemShut
  {NoStop}%
\end{thebibliography}%

\newpage
\clearpage


\section*{Supplementary material}

Chao-Ran Cai,[1] Zhi-Xi Wu,[1] Michael Z. Q. Chen,[2] Petter Holme,[3] and Jian-Yue Guan[1]
[1]*Institute of Computational Physics and Complex Systems,
Lanzhou University, Lanzhou, Gansu 730000, China*
[2]*Department of Mechanical Engineering, The University of Hong Kong, Pokfulam Road, Hong Kong*
[3]*Department of Energy Science, Sungkyunkwan University, Suwon 440-746, Korea*

In this supplemental file, we complement brief reviews on the popular analytical treatments of the SIR and SIS processes in heterogeneous networks (mainly developed in the last decade), the algorithm details of our stochastic simulations, and also the detailed theoretical analysis of the contact process in static networks.

### APPENDIX A: SIR EPIDEMIC PROCESS

In the classical SIR model, individuals can be in one of the three states, susceptible, infected, and recovered. Susceptible individuals become infected by being in contact with infected individuals at rate $r$ multiplying the number of susceptible-infected contacts, and infected individuals become recovered at rate $g$ and will be immune to the disease forever.

#### 1. SIR model in annealed networks—Heterogeneous mean-field theory (HMF)

The Heterogeneous Mean-Field theory is a degree-based mean-field theory. It is convenient to denote the relative density of infected, recovered and susceptible individuals of degree $k$ by $I_k$, $R_k$, and $S_k$, respectively (note the formulation here is not the same as that in the main paper). The HMF dynamical equations are given by (we set $\lambda = r/g$ and $g = 1$ without loss of generality)

$$\frac{d}{dt}S_k(t) = -\lambda k S_k(t) p_k(t), \tag{A.1}$$

$$\frac{d}{dt}R_k(t) = I_k(t), \tag{A.2}$$

where $p_k(t)$ is the probability of susceptible individual with given degree $k$ in contact with an infected individual at time $t$. For degree-uncorrelated random annealed networks, the probability that an edge pointing to an individual with degree $k'$ is equal to $k'P(k')/\langle k \rangle$ so that $p_k(t) = p(t)$ is independent of $k$, which can be read as

$$p(t) = \frac{1}{\langle k \rangle} \sum_k k P(k) I_k(t).$$

Combined with the initial conditions $I_k(0) \simeq 0$ and $S_k(0) \simeq 1$, Eq. (A.1) can be directly integrated,

$$S_k(t) = e^{-\lambda k \phi(t)},$$

where $\phi(t) = \int_0^t p(t)dt = \frac{1}{\langle k \rangle}\sum_k kP(k)R_k(t)$ is an auxiliary function.

Using the final state $\phi_\infty = \lim_{t\to\infty}\phi(t)$, $I_k(\infty) = 0$, and $R_k(\infty) = 1 - S_k(\infty)$, we have

$$R_\infty = \sum_k P(k)(1 - e^{-\lambda k \phi_\infty}). \tag{A.3}$$

We can calculate the time derivative of $\phi(t)$, $\frac{\phi(t)}{dt} = 1 - \phi(t) - \frac{1}{\langle k \rangle}\sum_k kP(k)e^{-\lambda k\phi(t)}$. Since $\lim_{t\to\infty}d\phi(t)/dt = 0$, we can obtain the following self-consistent equation for $\phi_\infty$,

$$\phi_\infty = 1 - \frac{1}{\langle k \rangle}\sum_k kP(k)e^{-\lambda k \phi_\infty}. \tag{A.4}$$

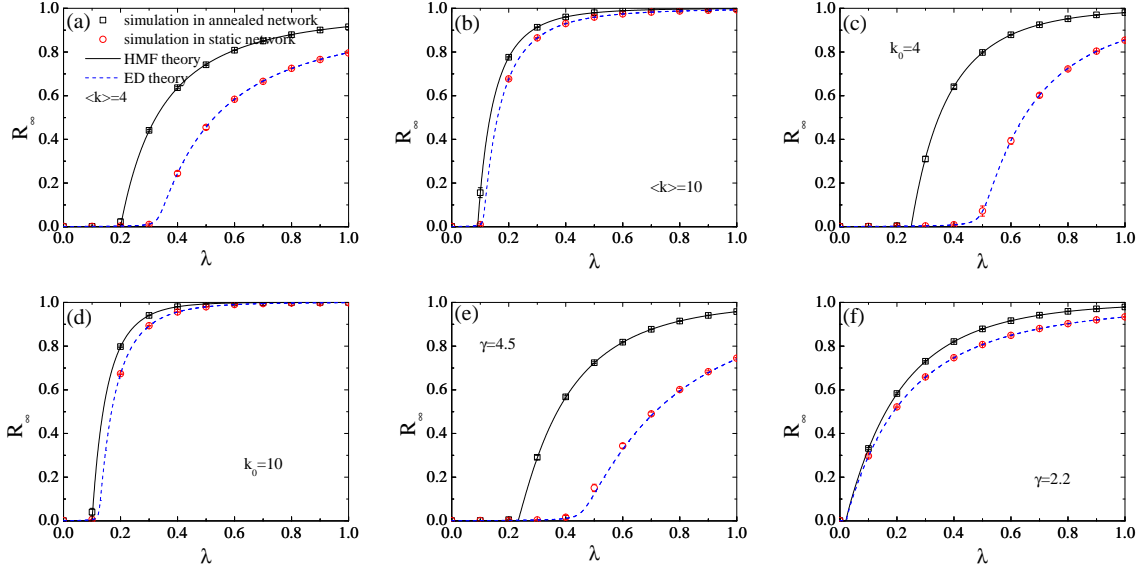

FIG. 1: (Color online) Epidemic prevalence $R_\infty$ in the SIR model are plotted as a function of the effective infection rate $\lambda$ in annealed and static networks. (a) and (b) ER random networks with average degree $\langle k \rangle = 4$ and 10; (c) and (d) RRGs with degree $k_0 = 4$ and 10; (e) and (f) SF networks with minimum degree 3 and $\gamma = 4.5$, 2.2. Simulations were implemented on networks with the total number of nodes $N = 10^5$ and the number of initial infected seeds $I_0 = 10^{-3}N$, and each data point is an average over at least 100 different epidemic outbreaks, performed on at least ten different network realizations.

To determine the final epidemic prevalence, we do as follows: i) solve Eq. (A.4) for $\phi_\infty$, ii) substitute $\phi_\infty$ into Eq. (A.3) to get the prevalence $R_\infty$. And then we can obtain the epidemic threshold from Eq. (A.4) by using the standard way [1],

$$\lambda_c^{\text{HMF,SIR}} = \frac{\langle k \rangle}{\langle k^2 \rangle}. \tag{A.5}$$

From Fig. 1, we can see that Eq. (A.5) predicted by the HMF theory is in well agreement with the simulations for the SIR process in annealed networks, and it serves as a good approximation for the SIR process in static RRGs and ER networks with large average degree, SF networks with $\gamma < 3$. Nevertheless, great discrepancies appear when the underlying static (homogeneous) networks have small average degree, or SF networks with $\gamma > 3$.

## 2. SIR model in static networks—Effective degree approach (ED)

The Effective Degree approach, considering explicitly the dynamic correlations between directly connected neighbors, is a higher-order degree-based mean-field approach which uses the mean-field approach in the calculation of susceptible neighbor to be infected.

The number of infected and susceptible individuals with $s$ susceptible neighbors and $i$ infected neighbors are represented by $I_{si}$ and $S_{si}$, respectively. The rate of new infections is $\sum_{k=0}^{k_{\max}} \sum_{j+l=k} \beta l S_{jl}$. These new infections cause their susceptible neighbors to change their effective degree at the rate $\sum_{k=0}^{k_{\max}} \sum_{j+l=k} j\beta l S_{jl}$. The total number of the susceptible neighbors of $S$ individuals is $\sum_{k=0}^{k_{\max}} \sum_{j+l=k} j S_{jl}$. So, the rate of one susceptible neighbor, belonging to an $S$ individual, to be infected is $\frac{\sum_{k=0}^{k_{\max}} \sum_{j+l=k} j\beta l S_{jl}}{\sum_{k=0}^{k_{\max}} \sum_{j+l=k} j S_{jl}}$. Similarly, the rate of one susceptible neighbor, belonging to an $I$ individual, to be infected is $\frac{\sum_{k=0}^{k_{\max}} \sum_{j+l=k} \beta l^2 S_{jl}}{\sum_{k=0}^{k_{\max}} \sum_{j+l=k} j I_{jl}}$.

The dynamic equations in the framework of ED approach are given by [2, 3]

$$\frac{dS_{si}}{dt} = -\beta i S_{si} + \gamma[(i+1)S_{s,i+1} - iS_{si}] + \frac{\sum_{k=0}^{k_{max}} \sum_{j+l=k} j\beta l S_{jl}}{\sum_{k=0}^{k_{max}} \sum_{j+l=k} j S_{jl}}[(s+1)S_{s+1,i-1} - sS_{si}], \tag{A.6}$$

$$\frac{dI_{si}}{dt} = \beta i S_{si} - \gamma I_{si} + \gamma[(i+1)I_{s,i+1} - iI_{si}] + \frac{\sum_{k=0}^{k_{max}} \sum_{j+l=k} \beta l^2 S_{jl}}{\sum_{k=0}^{k_{max}} \sum_{j+l=k} jI_{jl}}[(s+1)I_{s+1,i-1} - sI_{si}]. \quad (A.7)$$

The epidemic prevalence can be found by iterating the large sets of coupled Eq. (A.6) and (A.7). A lengthy but standard calculation [2] of the epidemic threshold gives

$$\lambda_c^{\text{ED,SIR}} = \frac{\langle k \rangle}{\langle k^2 \rangle - 2\langle k \rangle}.$$

Here it is worth pointing out that by using the ED approach, an exact expression for the epidemic threshold can be acquired clearly, but the calculation of prevalence is not accessible for a large network size.

### 3. Mapping the SIR model to the bond-percolation process

Define $T_{ij}$ the probability of individual $i$ transmitting disease to individual $j$, where $i$ is infected and $j$ is susceptible. There are two possible events occurred in $i$: it is cured to the recovered state with rate $g$, or it is in contact with the nearest neighbors and infect them with rate $k_i r$, where $k_i$ is the degree of $i$. So, the expected value of $T_{ij}$ can be written as

$$T_{ij} = \frac{g}{k_i r + g} \sum_{x=0}^{\infty} [\frac{(k_i - 1)r}{k_i r + g}]^x = \frac{g}{g + r} = \frac{1}{1 + \lambda},$$

where $x$ is the times of event occurred before individual $i$ becomes recovered and $\lambda = r/g$. We observe that $T_{ij}$ is a constant in the SIR epidemic process, so that we can map the SIR model to the occupation probability of percolation model. Since the critical occupation probability in random graphs is given by $p_c = \frac{\langle k \rangle}{\langle k^2 \rangle - \langle k \rangle}$ [4, 5]. So, by setting $T_{ij} = p_c$ we can easily obtain the epidemic threshold

$$\lambda_c^{percolation,SIR} = \frac{\langle k \rangle}{\langle k^2 \rangle - 2\langle k \rangle}.$$

By means of this approach, we can obtain an exact epidemic threshold. However, the effect of emerged loops in the case of a large number of infected is not negligible (especially when the network average degree is large), that is to say, we cannot approximate the random network to a tree. As a result, the mapping-to-percolation approach is not suitable for obtaining the epidemic prevalence.

## APPENDIX B: SIS EPIDEMIC PROCESS

In the SIS model, individuals can be in one of the two states, either susceptible or infected. Susceptible individuals become infected by in contact with infected individuals at rate $r$ multiplying the number of susceptible-infected contacts, and infected individuals get cured to be susceptible again at rate $g$.

### 1. SIS model in annealed networks—Heterogeneous mean-field theory

In the steady state of the SIS epidemic process, we have

$$-I_k + \lambda k(1 - I_k)\phi = 0, \quad (B.1)$$

where $\phi = \frac{1}{\langle k \rangle} \sum_k kP(k)I_k$. In the steady state, we can obtain a self-consistent equation for $\phi$,

$$\phi = \frac{1}{\langle k \rangle} \sum_k kP(k) \frac{\lambda k \phi}{1 + \lambda k \phi}. \quad (B.2)$$

We can calculate the epidemic prevalence as follows: i) solve Eq. (B.2) for $\phi$, ii) substitute $\phi$ to Eq. (B.1) to obtain $I_k$, iii) yield the epidemic prevalence $\rho = \sum_k I_k P(k)$. By implementing the standard manipulation [1], we can get the epidemic threshold as

$$\lambda_c^{\text{HMF,SIS}} = \frac{\langle k \rangle}{\langle k^2 \rangle}.$$

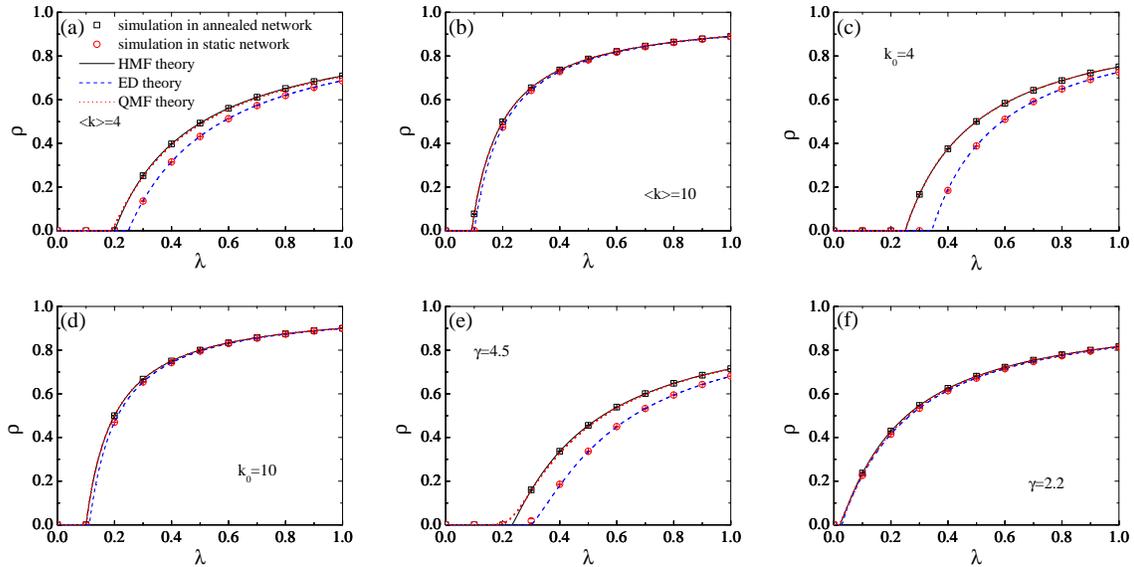

FIG. 2: (Color online) The epidemic prevalence $\rho$ in the SIS process is plotted as a function of the effective infection rate $\lambda$ in annealed and static networks. (a) and (b) ER random networks with average degree $\langle k \rangle = 4$ and 10; (c) and (d) RRGs with degree $k_0 = 4$ and 10; (e) and (f) SF networks with minimum degree 3 and $\gamma = 4.5$, 2.2. Simulations were implemented on networks with the total number of nodes $N = 10^5$ and the number of initial infected seeds $I_0 = 1\text{\textperthousand} \times N$. Each data point is an average over at least 100 different epidemic outbreaks, performed on at least 10 different network realizations.

From Fig. 2, we can see that HMF theory gives rise to perfect results for the SIS process in annealed networks, and it serves as a good candidate to analyze the SIS process in static RRGs and ER networks with large average degrees, SF networks with $\gamma < 3$. Similar to the case of SIR, great discrepancies appear when the underlying static (homogeneous) networks have small average degrees, or SF networks with $\gamma > 3$.

### 2. SIS model in static networks—Quenched mean-field theory (QMF)

The Quenched Mean-Field theory is an individual-based mean-field theory. It takes into account full network structure correlation and ignores the dynamic correlation. The infection probability of an individual $i$ is represented by $I_i$. The QMF dynamical equations are given by

$$\frac{dI_i}{dt} = -I_i + \lambda S_i \sum_j I_j A_{ij}, \tag{B.3}$$

where $A_{ij}$ is the adjacency matrix with value $A_{ij} = 1$ if individuals $i$ and $j$ are connected, and zero otherwise. Here, we can obtain the epidemic prevalence by iterating the large sets of coupled Eq. (B.3) to the steady state.

In the stationary state, we obtain the linearized equation

$$\frac{dI_i}{dt} = \sum_j (-\delta_{ij} + \lambda A_{ij}) I_j,$$

implying that the epidemic threshold is given by [6]

$$\lambda_c^{\text{QMF,SIS}} = \frac{1}{\Lambda_1},$$

where $\Lambda_1$ is the largest eigenvalue of the adjacency matrix of the underlying interaction network.

By using this approach, the calculation of the epidemic prevalence is also greatly limited when the network size is becoming large.

### 3. SIS model in static networks—Effective degree approach

The dynamical equations derived by the ED approach for the SIS epidemic process are given by

$$\frac{dS_{si}}{dt} = -\beta i S_{si} + \gamma I_{si} + \gamma[(i+1)S_{s-1,i+1} - iS_{si}] + \frac{\sum_{k=0}^{k_{max}} \sum_{j+l=k} j\beta l S_{jl}}{\sum_{k=0}^{k_{max}} \sum_{j+l=k} j S_{jl}}[(s+1)S_{s+1,i-1} - sS_{si}], \quad \text{(B.4)}$$

$$\frac{dI_{si}}{dt} = \beta i S_{si} - \gamma I_{si} + \gamma[(i+1)I_{s-1,i+1} - iI_{si}] + \frac{\sum_{k=0}^{k_{max}} \sum_{j+l=k} \beta l^2 S_{jl}}{\sum_{k=0}^{k_{max}} \sum_{j+l=k} j I_{jl}}[(s+1)I_{s+1,i-1} - sI_{si}]. \quad \text{(B.5)}$$

The epidemic prevalence can be yielded by iterating the large sets of coupled Eq. (B.4) and (B.5) to the stationary state. One can also prove that epidemic threshold by the ED method satisfies the following relation [2]

$$\lambda_c^{ED,SIS} < \lambda_c^{ED,SIR} = \frac{\langle k \rangle}{\langle k^2 \rangle - 2\langle k \rangle}.$$

### APPENDIX C: MOTIVATIONS

In annealed networks (where dynamic correlation is totally ignored), the epidemic threshold is the same for the SIS and SIR epidemic processes. The HMF theory can provide a simple analytical solution for the epidemic prevalence, and also successfully predict the epidemic threshold for both epidemic processes. In the static networks, where the dynamic correlation is present and cannot be overlooked, the epidemic threshold of the SIS process is smaller than that of the SIR process. The efficiency of the HMF, as an approximation, is limited, for homogeneous networks with small average degree and for SF networks with $\gamma > 3$. And the QMF and ED approaches, formulated by on a large set of coupled ordinary differential equations, cannot provide an explicit analytical solution of the epidemic prevalence for both SIS and SIR epidemic processes in networks with large size. Moreover, the prediction of epidemic threshold by the QMF does not match well with the Monte Carlo simulations for SIS epidemic process in both RRGs and ER networks.

To resolve the aforementioned problems, in the main paper we study how the presence of dynamic correlation affects the epidemic dynamics. We obtain the succinct analytical solution for the SIS epidemic process and improve the prediction accuracy of the epidemic threshold as compared to the QMF theory.

### APPENDIX D: DETAILS FOR THE STOCHASTIC SIMULATIONS AND DERIVATIONS

#### 1. Generation of static, uncorrelated networks with arbitrary degree distribution

We use the configuration model algorithm presented in [7] to generate static, uncorrelated networks. In particular, for any network with a specified degree distribution, the configuration model can be expressed with a stub-connection algorithm [7]: For each node we first assign it a degree $k_i$ (whose value is restricted in the range of $[k_{\min}, \sqrt{N}]$, where $k_{\min}$ is the minimum degree of the node in the network and $\sqrt{N}$ the network size) according to the prescribed degree distribution, and then create a set of $k_i$ stubs that represent each of these edges with only a single tail connected to a node; We repeat this for all nodes and then combine these stubs into a master set; This set is then randomly divided in half, and a stub from each subset is matched to one from the other subset, forming a complete edge. If there is an uneven number of stubs, a random individual is given an extra stub. Self connections or duplicate edges between nodes are not allowed in the generation process.

#### 2. Simulation procedure of the epidemic prevalence in static networks

At any time $t$, we calculate each individual's transition rate $\eta_i(t)$. The rate for any susceptible individual becoming infected is $\eta_i(t) = r \times k_{inf}$ and $k_{inf}$ is the number of infected neighbors of the focal individual. The rate for any infected individual getting cured is $\eta_i(t) = g$. Summing up all of them, we yield the total transition rate $\omega(t) = \sum_i \eta_i(t)$. With this value in hand, the time at which the next transition event occurs is $t' = t + dt$, where $dt = 1/\omega(t)$. The individual whose state is chosen to change at time $t'$ is sampled with a probability proportional to $\eta_i(t)$. That is, a

uniform random number $u \in [0, 1)$ is generated and if $\sum_{j=1}^{k-1} \eta_j(t)/\omega(t) < u < \sum_{j=1}^{k} \eta_j(t)/\omega(t)$, then individual $k$ is chosen to change state. The whole process is iterated until the population reach to a stationary state, where either an absorbing state of all susceptible individuals arises or an endemic equilibrium is arrived (i.e., the number of infected individuals fluctuates stably in the long time limit).

### 3. Simulation procedure of the epidemic prevalence in annealed networks

At any time $t$, we know the selected rate of each infected individual $i$ is $\eta_i(t) = g + rk_i$ and $k_i$ is the degree of individual $i$. Summing up all of them, we yield the total transition rate $\omega(t) = \sum_{i \in I} \eta_i(t)$. And time is incremented by $dt = 1/\omega(t)$. The infected individual who is selected at time $t + dt$ is sampled with a probability proportional to $\eta_i(t)$. Then, with probability $g/\eta_j(t)$, the selected individual $j$ is assumed be cured. With probability $rk_j/\eta_j(t)$, an infection attempt is performed in two steps: (i) another individual $j'$ who will be contacted by $j$ is selected with probability proportional to its degree from the whole population excluding $j$ itself. (ii) if the individual $j'$ is infected nothing happens, else it becomes infected. The whole process is iterated until the population reach a steady state.

### 4. Simulation procedure of the epidemic threshold in static networks

The simulations were performed using the quasi-stationary (QS) method [8] which permits to overcome the difficulties intrinsic to the simulations of finite systems with absorbing states. This procedure is implemented by replacing the absorbing state, every time the system tries to visit it, with an active configuration randomly taken from the history of the simulation. For this task, a list of $M$ active configurations, corresponding to states previously visited by the dynamics, is stored and constantly updated. An update consists of randomly choosing a configuration in the list and replacing it by the present active configuration with a probability $p_r \Delta t$. After a relaxation time $t_r$, the QS quantities are determined during an averaging time $t_a$. During the averaging time, the QS probability $\bar{P}(n)$ that the system has $n$ infected individuals is computed. From the distribution $\bar{P}(n)$, the moments of the activity distribution can be computed as $\langle \rho^k \rangle = \sum_n (n/N)^k \bar{P}(n)$ and the susceptibility defined as $\chi = N(\langle \rho^2 \rangle - \langle \rho \rangle^2)/\langle \rho \rangle$. When plotted as a function of $\lambda$ in a system of fixed size $N$, the susceptibility $\chi$ exhibits a maximum at a value $\lambda_p(N)$, which determines the critical effective infection rate for finite system size $N$ [9, 10]. The values of the QS parameters used in the present simulations were $M = 100$, $p_r = 0.02$, $t_r = 10^5$, and $t_a$ are varied from $3 \times 10^5$ to $10^7$ depending on $N$ and $\lambda$.

### 5. Derivation of Eq. (7) in the main text

The left side of Eq. (7) in the main paper can be calculated as follows,

$$\sum_k \sum_j S_{k,j} j^2 r = \sum_k r \sum_j S_k (1-p_k)^{k-j} p_k^j \binom{k}{j} j^2 = \sum_k rkp_k S_k[(k-1)p_k + 1] = \sum_k gI_k[(k-1)p_k + 1],$$

where we have employed Eq. (4) in the main text.

For the right side of Eq. (6) in the main paper, we notice that $\sum_j I_{k,j} j$ is the total number of infected-infected pairs with degree $k$. From the HMF theory, we know that the total number of susceptible-infected pair with degree $k$ is $\frac{I_k g}{r}$. And considering a balance condition that the total number of susceptible-infected pair must be equal to the total number of infected-susceptible pair in static networks in any case, we get

$$\sum_k \sum_j I_{k,j} jg = \sum_k g \left[ (\sum_j I_{k,j} k) - I_k \frac{g}{r} \right] = \sum_k gI_k[k - \frac{g}{r}].$$

Then, we can obtain Eq. (7) in the main paper:

$$g \sum_k I_k \left[ (k-1)p_k + 1 - k + \frac{g}{r} \right] = 0.$$

## 6. Proof of $q_{k_0} > p_{k_0}$ for SIS process on static random regular graphs

For SIS dynamics taking place upon static random regular graphs, we are able to give a strict proof of the inequality $q_{k_0} > p_{k_0}$. Denote the number of susceptible-infected pairs, susceptible-susceptible pairs, and infected-infected pairs by $\prod_{SI}$, $\prod_{SS}$ and $\prod_{SI}$, respectively. In the stationary (endemic) state, the total number of susceptible-infected pairs must be equal to that of infected-susceptible pairs, according to which we have $\prod_{IS} = \prod_{SI} = Ig/r$. With this value at hand, we can obtain the total number of infected-infected and susceptible-susceptible pairs in the steady state, $\prod_{II} = k_0 I - Ig/r$ and $\prod_{SS} = k_0(N-I) - Ig/r$, respectively. According the definition of $q_{k_0}$ and $p_{k_0}$ (the probabilities of reaching an arbitrary infected individual by following a randomly chosen edge from infected and susceptible individuals of degree $k_0$), we have

$$q_{k_0} = \frac{\prod_{II}}{\prod_{II} + \prod_{IS}} = 1 - \frac{1}{k_0 \lambda}, \quad p_{k_0} = \frac{\prod_{SI}}{\prod_{SI} + \prod_{SS}} = \frac{1}{k_0 \lambda} \frac{\rho}{1-\rho}, \tag{D.1}$$

where $\lambda = r/g$ (the effective infection rate) and $\rho = I/N$ (the epidemic prevalence in the stationary endemic state). By considering the relationship of $\frac{1}{1-\rho} = 1 + k_0(\lambda - \frac{1}{k_0-1})$ [Eq.(14) of main text], we can yield

$$q_{k_0} - p_{k_0} = 1 - \frac{1}{k_0 \lambda} - \frac{1}{k_0 \lambda} \frac{\rho}{1-\rho} = 1 - \frac{1}{k_0 \lambda} \frac{1}{1-\rho} = \frac{1}{k_0 \lambda} \frac{1}{k_0 - 1} > 0. \tag{D.2}$$

For SIS model on (static) heterogenous networks, we were not yet able to make a rigorous proof for the inequality $q_k > p_k$. But one might understand this point as follows: For a degree-$k$ susceptible, the neighbors could be composed of both susceptible and infected individuals, or just of susceptible individuals. However, for any infected individual, it must have been infected by one of its infected neighbors before, so that it is likely that its neighborhood has some infected individuals. Thus, assuming the relationship $q_k > p_k$ seems acceptable.

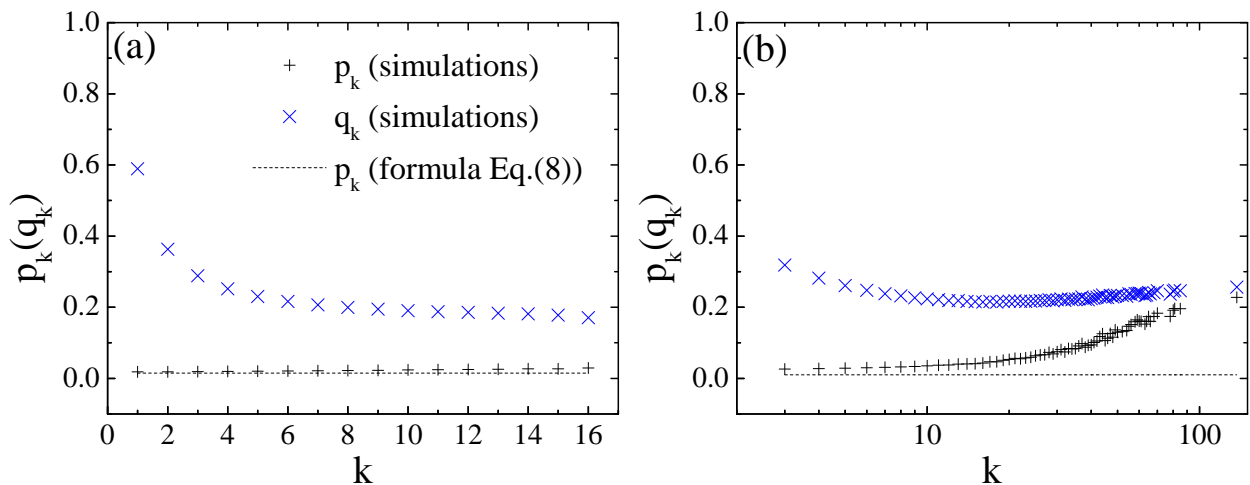

FIG. 3: (Color online) The probability $p_k$ ($q_k$) of a susceptible (infected) individual with degree $k$ contact an arbitrary infected neighbor is shown for various node degree. Parameters: (a) Erdős-Rényi random networks with average degree 4 and $\lambda = 0.255$; (b) scale-free network with minimum degree 3, $\gamma = 4.5$, and $\lambda = 0.305$. Simulations were implemented on networks with the total number of nodes $N = 10^6$ and a fraction $10^{-3}$ of them are seeds. Each data point is an average over at least 1000 independent simulation runs per network realization.

## 7. Numerical validation of Eqs. (2) and (8) in the main text

Since the assumption of Eq. (8) is crucial in the derivation of the epidemic prevalence and threshold, we provide a numerical validation of it in Fig. 3, where we calculate the frequencies of $p_k$ and $q_k$ during the endemic state for both homogeneous and heterogeneous networks. We observe (Fig. 3) that $p_k$ changes very slowly with $k$. For homogeneous networks [Fig. 3(a)], $p_k$ is close to constant, which confirms our hypothesis of Eq. (8). For heterogeneous networks

TABLE I: All possible situations in an interval time $dt$ of the contact process in static RRGs.

| situation | probability | $\Delta a$ | $\Delta b$ |
|---|---|---|---|
| $S_{k_0,j} \to I_{k_0,j}$ | $\frac{S_{k_0,j}(t)jr\frac{1}{k_0}}{a'+b'}$ | $+g$ | $[-jr+r(k_0-j)]\frac{1}{k_0}$ |
| $I_{k_0,j} \to S_{k_0,j}$ | $\frac{I_{k_0,j}(t)g}{a'+b'}$ | $-g$ | $[+jr-r(k_0-j)]\frac{1}{k_0}$ |

[Fig. 3(b)], a deviation of $p_k$ obtained from the simulations and our assumption appears for very large degree class. (Note that in scale-free networks, the vast majority of nodes belong to the small degree class.) Moreover, we also notice that $q_k$ is indeed always greater than $p_k$, corroborating our conjecture Eq. (2). Taken together, we argue that Eqs. (2) and (8) capture the most important parts of the dynamic correlations, neglecting only minor ones.

### 8. Extended to the contact process (CP) in static random regular graphs

The CP is the prototypical model of the directed percolation class. The time evolution of the CP in static RRGs runs as follows: At any time $t$, an infected individual $i$ is chosen at random. With probability $\frac{g}{g+r}$, the individual $i$ becomes cured. With probability $\frac{r}{g+r}$, an infection attempt is performed in two steps: (i) an individual $j$, one nearest neighbor of the individual $i$, is randomly chosen. (ii) if the individual $j$ is infected nothing happens, else it becomes infected. The time is incremented by $dt = 1/[(g+r)n(t)]$, where $n(t)$ is the number of infected individual at the beginning of the time $t$. The whole process is iterated until the population reach to a steady state.

In CP, the total recovery and transmission rates of the whole population are $a' = \sum_j I_{k_0,j}g$ and $b' = \sum_j S_{k_0,j}jr\frac{1}{k_0}$, respectively. And now, the Eq. (4) in the main paper will be replaced by $p = \frac{g}{r}\frac{I_{k_0}}{S_{k_0}}$. Considering the Table. I, we can obtain

$$\frac{r}{k_0}\sum_j S_{k_0,j}j^2 = g\sum_j I_{k_0,j}j. \tag{D.3}$$

We can calculate the left side of Eq. (D.3),

$$\begin{aligned}\frac{r}{k_0}\sum_j S_{k_0,j}j^2 &= \frac{r}{k_0}\sum_j S_{k_0}(1-p)^{k_0-j}p^j\binom{k_0}{j}j^2 \\ &= \frac{1}{k_0}rk_0 p S_{k_0}\left[(k_0-1)p+1\right] \\ &= gI_{k_0}\left[(k_0-1)p+1\right].\end{aligned}$$

We notice that $\sum_j I_{k_0,j}j$ is the total number of infected-infected pairs. We know that the total number of infected-susceptible pairs is $\frac{I_{k_0}g}{r}k_0$. Thus, the right side of Eq. (D.3) can be written as

$$\begin{aligned}g\sum_j I_{k_0,j}j &= g\left[I_{k_0}k_0 - I_{k_0}\frac{gk_0}{r}\right] \\ &= gI_{k_0}\left[k_0 - k_0\frac{g}{r}\right],\end{aligned}$$

Then, we can yield

$$gI_{k_0}\left[(k_0-1)p+1-k_0+k_0\frac{g}{r}\right] = 0.$$

Substituting $p = \frac{g}{r}\frac{I_{k_0}}{S_{k_0}}$ and $N_{k_0} = S_{k_0} + I_{k_0}$, we can obtain

$$\rho = \frac{I_{k_0}}{N_{k_0}} = 1 - \frac{1}{1+\lambda-\frac{k_0}{k_0-1}}. \tag{D.4}$$

We compare the results from theoretical estimations and those from Monte Carlo simulations in Fig. 4, and observe that they match quite well with each other.

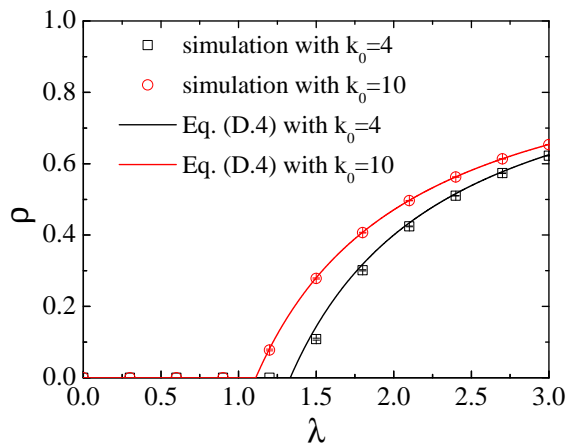

FIG. 4: (Color online) The epidemic prevalence are plotted as a function of the effective infective rate in static RRGs with degree $k_0 = 4$ and 10 on contact process.

# APPENDIX E: BRIEF REMARKS ON SEVERAL CLASSICAL THEORETICAL METHODS IN SOLVING SIMPLE EPIDEMIC SPREADING MODELS

In this last part, we would like to give some remarks on the three main stream methods, say, the HMF, QMF and ED (mostly developed in the last 15 years), in addressing simple epidemic spreading models (such as SIS and SIR) from a theoretical point of view:

1) The HMF theory ignores the dynamic and network structural correlations totally, and can provide explicit expressions for the epidemic threshold and prevalence. Although the theoretical predictions (of both epidemic prevalence and threshold) obtained from the HMF theory match quite well with the simulation results in annealed networks, its applicability in the case of static networks is largely limited, due to its overlook of the dynamic correlation.

2) The QMF theory considers full network-structural correlations, but also completely neglects the dynamic correlation. The epidemic threshold predicted from the QMF theory is better than that from the HMF theory. However, the calculation of the epidemic prevalence depends on a large set of coupled ordinary differential equations, which becomes not suitable for large-size networks. In addition, the calculated epidemic prevalence does not match well with the simulation results in the case of static networks.

3) The ED method takes into account of the dynamic correlation and structural correlation between directly connected neighbors, and omits necessarily high-order network structural correlation. Both the estimated epidemic prevalence and threshold from ED method fit the simulation results. However, as in the case of QMF, it also depends on a large set of coupled ordinary differential equations, which becomes not suitable for large-size networks. Besides, one cannot find explicit forms of the epidemic prevalence and threshold by using ED.

As presented in the main text, the combination of HMF and ED methods can complement each other, through which we are able to, on one hand, include dynamic correlation and necessary structural correlation to get more accurate predictions (than in previous works) of both the epidemic prevalence and threshold, and on the other hand obtain explicit expressions for these two quantities.

[1] R. Pastor-Satorras and A. Vespignani, Phys. Rev. Lett. **86**, 3200 (2001), URL http://link.aps.org/doi/10.1103/PhysRevLett.86.3200.
[2] J. Lindquist, J. Ma, P. Driessche, and F. Willeboordse, J. Math. Biol. **62**, 143 (2011).
[3] C.-R. Cai, Z.-X. Wu, and J.-Y. Guan, Phys. Rev. E **90**, 052803 (2014), URL http://link.aps.org/doi/10.1103/PhysRevE.90.052803.
[4] D. S. Callaway, M. E. J. Newman, S. H. Strogatz, and D. J. Watts, Phys. Rev. Lett. **85**, 5468 (2000), URL http://link.aps.org/doi/10.1103/PhysRevLett.85.5468.
[5] R. Cohen, K. Erez, D. ben Avraham, and S. Havlin, Phys. Rev. Lett. **85**, 4626 (2000), URL http://link.aps.org/doi/10.1103/PhysRevLett.85.4626.

[6] R. Pastor-Satorras, C. Castellano, P. Van Mieghem, and A. Vespignani, Rev. Mod. Phys. **87**, 925 (2015), URL http://link.aps.org/doi/10.1103/RevModPhys.87.925.
[7] M. Catanzaro, M. Boguñá, and R. Pastor-Satorras, Phys. Rev. E **71**, 027103 (2005), URL http://link.aps.org/doi/10.1103/PhysRevE.71.027103.
[8] M. M. de Oliveira and R. Dickman, Phys. Rev. E **71**, 016129 (2005), URL http://link.aps.org/doi/10.1103/PhysRevE.71.016129.
[9] M. Boguñá, C. Castellano, and R. Pastor-Satorras, Phys. Rev. Lett. **111**, 068701 (2013), URL http://link.aps.org/doi/10.1103/PhysRevLett.111.068701.
[10] S. C. Ferreira, C. Castellano, and R. Pastor-Satorras, Phys. Rev. E **86**, 041125 (2012), URL http://link.aps.org/doi/10.1103/PhysRevE.86.041125.


\end{document}